\documentclass[]{sn-jnl}

\usepackage{anyfontsize}
\usepackage{graphicx}
\usepackage{multirow}
\usepackage{amsmath,amssymb,amsfonts}
\usepackage{amsthm}
\usepackage{mathrsfs}
\usepackage{siunitx}
\usepackage[title]{appendix}
\usepackage{xcolor}
\usepackage{textcomp}
\usepackage{manyfoot}
\usepackage{booktabs}
\usepackage{algorithm}
\usepackage{algorithmicx}
\usepackage{algpseudocode}
\usepackage{listings}
\usepackage{soul}
\usepackage{svg}
\usepackage{adjustbox}
\usepackage{float}
\usepackage{graphicx}
\usepackage{caption}
\usepackage{subcaption}
\usepackage{threeparttable}
\usepackage{xargs}
\usepackage{tabularx}
\usepackage{comment,lineno,acronym,blindtext}
\usepackage{setspace}

\usepackage[style=numeric-comp, sorting=none, firstinits, isbn=false,
eprint=false, doi=true, url=false, arxiv=abs, backend=biber]{biblatex}

\def\BibTeX{{\rm B\kern-.05em{\sc i\kern-.025em b}\kern-.08em
    T\kern-.1667em\lower.7ex\hbox{E}\kern-.125emX}}

\usepackage{hyperref}
\hypersetup{colorlinks=True,linkcolor={pomgrenate},citecolor={blue},urlcolor={blue}}
\DeclareSourcemap{
  \maps{
    \map{
      \step[fieldset=month, null]
      \step[fieldset=publisher,null]
      \step[fieldset=address, null]
      \step[fieldset=location, null]
      \step[fieldset=url, null]
      \step[fieldset=isbn, null]
      \step[fieldset=number, null]
      \step[fieldset=note, null]
      \step[fieldset=urldate, null]
      \step[fieldset=language,null]
      \step[fieldset=issn, null]
      \step[fieldset=volume, null]
      \step[fieldset=editor, null]
    }
  }
}
\DeclareGraphicsExtensions{.jpg,.pdf,.mps,.png}
\graphicspath{{figures} {img/} {imglocal/} {./}} 

\definecolor{pomgrenate}{RGB}{192, 57, 43}
\definecolor{midnightblue}{RGB}{44, 62, 80}
\definecolor{greensea}{RGB}{22, 160, 133}
\definecolor{pumpkin}{RGB}{211, 84, 0}
\definecolor{wisteria}{RGB}{142, 68, 173}
\definecolor{orange}{RGB}{243, 156, 18}
\definecolor{green}{RGB}{168, 205, 159}
\definecolor{jade}{RGB}{1, 163, 164}
\definecolor{belize}{RGB}{41,128,185}

\acrodef{EMG}[EMG]{electromyography}
\acrodef{iEMG}[iEMG]{intramuscular electromyography}
\acrodef{sEMG}[sEMG]{surface electromyography}
\acrodef{HD-iEMG}[HD-iEMG]{high-density intramuscular electromyography}
\acrodef{HD-sEMG}[HD-sEMG]{high-density surface electromyography}
\acrodef{LR}[LR]{linear regression}
\acrodef{MLP}[MLP]{multi-layer perceptron}
\acrodef{SNN}[SNN]{Spiking Neural Network}
\acrodef{DoF}[DoF]{Degrees-of-Freedom}
\acrodef{MUAP}[MUAP]{motor unit action potential}
\acrodef{MVC}[MVC]{maximal voluntary contraction}
\acrodef{RMS}[RMS]{root mean square}
\acrodef{MAV}[MAV]{mean absolute value}
\acrodef{RMSE}[RMSE]{root mean squared error}
\acrodef{MAE}[MAE]{mean absolute error}
\acrodef{R2}[R$^2$]{coefficient of determination}
\acrodef{LIF}[LIF]{leaky integrate-and-fire}
\acrodef{LI}[LI]{leaky integrator}
\acrodef{CKC}[CKC]{convolutive kernel compensation}
\acrodef{iBMI}[iBMI]{implantable brain-machine interface}
\acrodef{MEA}[MEA]{microelectrode array}
\begin{document}
\title[Article Title]{Spiking Neural Network Decoders of Finger Forces from High-Density Intramuscular Microelectrode Arrays}

\author*[1]{\fnm{Farah} \sur{Baracat}}\email{fbarac@ini.uzh.ch}
\author[2]{\fnm{Agnese} \sur{Grison}}\email{agnese.grison16@imperial.ac.uk}
\author[2]{\fnm{Dario} \sur{Farina}}\email{d.farina@imperial.ac.uk}
\author[1]{\fnm{Giacomo} \sur{Indiveri}}\email{giacomo@ini.uzh.ch}
\author*[1]{\fnm{Elisa} \sur{Donati}}\email{elisa@ini.uzh.ch}

\affil*[1]{\orgdiv{Institute of Neuroinformatics}, \orgname{University of Zurich and ETH Zurich}, \orgaddress{\city{Zurich}, \country{Switzerland}}}

\affil[2]{\orgdiv{Department of Bioengineering}, \orgname{Imperial College London}, \orgaddress{ \city{London}, \country{United Kingdom}}}
\abstract{Restoring naturalistic finger control in assistive technologies requires the continuous decoding of motor intent with high accuracy, efficiency, and robustness. Here, we present a spike-based decoding framework that integrates spiking neural networks (SNNs) with motor unit activity extracted from high-density intramuscular microelectrode arrays. We demonstrate simultaneous and proportional decoding of individual finger forces from motor unit spike trains during isometric contractions at 15\% of maximum voluntary contraction using SNNs. We systematically evaluated alternative SNN decoder configurations and compared two possible input modalities: physiologically grounded motor unit spike trains and spike-encoded intramuscular EMG signals. Through this comparison, we quantified trade-offs between decoding accuracy, memory footprint, and robustness to input errors. The results showed that shallow SNNs can reliably decode finger-level motor intent with competitive accuracy and minimal latency, while operating with reduced memory requirements and without the need for external preprocessing buffers. This work provides a practical blueprint for integrating SNNs into finger-level force decoding systems, demonstrating how the choice of input representation can be strategically tailored to meet application-specific requirements for accuracy, robustness, and memory efficiency.}

\maketitle

\section{Introduction}
\label{sec:introduction}
Myoelectric systems interface with the peripheral nervous system by recording the electrical activity of skeletal muscles through \ac{EMG}, using either non-invasive surface electrodes or minimally invasive intramuscular electrodes~\cite{parker_signal_1977,wolczowski_humanmachine_2010,tam_human_2019}. Although \ac{EMG} signals do not directly capture neural activity, they reflect the electrical activity of muscle fibers innervated by motor neurons, providing an indirect measure of the cumulative output of spinal motor circuits~\cite{day_experimental_2001}. Even so, \ac{EMG}-based systems enable the inference of user intent and have been widely employed to control external devices such as neuroprostheses and exoskeletons~\cite{parker_myoelectric_1986, fu_myoelectric_2022, xiong_robotic_2024}, as well as for applications in virtual and augmented environments~\cite{lin_vr-based_2023}.

Conventional myoelectric decoding approaches typically rely on specific features extracted from \ac{EMG} signals, such as time-domain statistics or frequency components to estimate the motor intent~\cite{jiang_myoelectric_2012, boostani_evaluation_2003, roche_prosthetic_2014}. These methods have proven effective in classification tasks for identifying discrete movement classes, as well as in proportional and simultaneous control of multiple \acp{DoF}~\cite{jiang_emg-based_2012,young_classification_2013,  hahne_linear_2014}. However, reliance on global \ac{EMG} features presents several key limitations: signal features are often sensitive to electrode placement, skin-electrode impedance variations, and inter-subject differences. In addition they do not directly capture the underlying neural commands originating from spinal motor neurons~\cite{del_vecchio_tutorial_2020, keenan_influence_2005, thompson_robust_2018}. As such, these global features offer only an indirect and coarse estimate of the intended neural drive to the muscles.

A promising alternative leverages novel decomposition techniques to extract the precise discharge timings of individual motor units --the smallest functional elements of the motor system-- which provide a direct, physiologically grounded, and temporally accurate representation of the descending motor drive~\cite{gazzoni_new_2004,holobar_multichannel_2007,farina_extraction_2014}. Recent studies have demonstrated that decoding motor intent from motor unit activity outperforms traditional approaches based on global \ac{EMG} features, particularly in tasks demanding high-resolution proportional control of individual fingers~\cite{kapelner_predicting_2019,xu_estimation_2021, xia_simultaneous_2025}.

Despite their demonstrated performance gains, most motor unit-based decoding techniques make use of conventional machine learning models that are inherently designed for continuous or rate-based inputs, and are therefore ill-suited for discrete spiking data. To overcome this representational mismatch, motor unit spike events are typically aggregated into fixed-duration bins to compute discharge rates, converting the spiking signal back into a continuous format compatible with standard decoders~\cite{kapelner_predicting_2019, xia_simultaneous_2025}. However, this intermediate step introduces latency, and adds memory overhead due to the need for spike buffering and rate computation, motivating the development of alternative methods inherently suited to spike-based data processing.

In this work we propose a spike-based processing method where spike trains are streamed directly into \acp{SNN}, a class of neural networks optimally suited for real-time, event-driven computation~\cite{Maass_Bishop98,Gerstner_Kistler02}. Unlike conventional artificial neural networks, \acp{SNN} operate asynchronously and process inputs as discrete, time-resolved spike events, making them ideal for neuromorphic hardware implementations~\cite{Mead23,Donati_Indiveri23}. Within this framework, neurons and synapses act concurrently as computational and memory units, integrating input spikes and maintaining internal states via biologically inspired temporal dynamics. This tight coupling between processing and memory eliminates the need for spike binning or intermediate buffering, substantially reducing latency and computational complexity in real-time motor decoding applications~\cite{Indiveri_Sandamirskaya19}.

Despite the natural compatibility between motor unit spike trains and the event-driven nature of \acp{SNN}, their adoption in this domain remains largely unexplored. To date, only a single study has demonstrated this potential, focusing on gesture classification using \ac{HD-sEMG} and motor unit spike trains decomposed via the convolutive kernel compensation algorithm from surface EMG~\cite{holobar_multichannel_2007, tanzarella_neuromorphic_2023}.

In contrast, here, we adopt a systematic approach to the problem of decoding motor intent from intramuscular implanted \acp{MEA}, explicitly targeting real-time, resource-constrained applications. We compare classical and neuromorphic decoding pipelines by leveraging three key advancements. First, we demonstrate for the first time the proportional and simultaneous decoding of individual finger forces from motor unit spike trains identified via high-density intramuscular signals using \acp{MEA}. Second, we exploit recent advances in \ac{HD-iEMG} decomposition, which enable the identification of large motor unit populations and have previously shown promising results in classification tasks~\cite{muceli_accurate_2015, grison_particle_2024, grison_intramuscular_2024}. Third, we systematically evaluate the performance and memory footprint of two shallow \ac{SNN} architectures, offering a path toward low-power, embedded neuromorphic hardware implementations. Our decoding framework moves beyond gesture classification and toward fine-grained, temporally precise, and intuitive control.

 While motor unit activity provides a direct and physiologically meaningful input to \acp{SNN}, the decomposition algorithms are computationally demanding and memory intensive, due in part to the need for storing separation matrices~\cite{rossato_i-spin_2024, barsakcioglu_control_2021}. To overcome this challenge, we explore and evaluate a complementary event-based strategy that bypasses decomposition altogether by encoding filtered \ac{HD-iEMG} signals into spike trains using \ac{LIF} neurons~\cite{baracat_decoding_2024, baracat_heterogeneous_2025}.  Although such representations lacks the physiological equivalence of individual motor units, they offer a lightweight, hardware-compatible alternative that retains sufficient task-relevant information and incurs no additional memory cost. 
 
By comparing these decoding strategies, we expose the fundamental tradeoffs between physiologically meaningful inputs and implementation efficiency, providing an initial basis for selecting decoder architectures based on the specific constraints and objectives of the target application.

\section{Results}\label{sec:results}
\begin{figure}[ht!]
    \centering
    \begin{minipage}[b]{0.42\textwidth}
        \begin{adjustbox}{valign=b}
        \begin{subfigure}[b]{\linewidth}
            \centering
            \includegraphics[width=\linewidth]{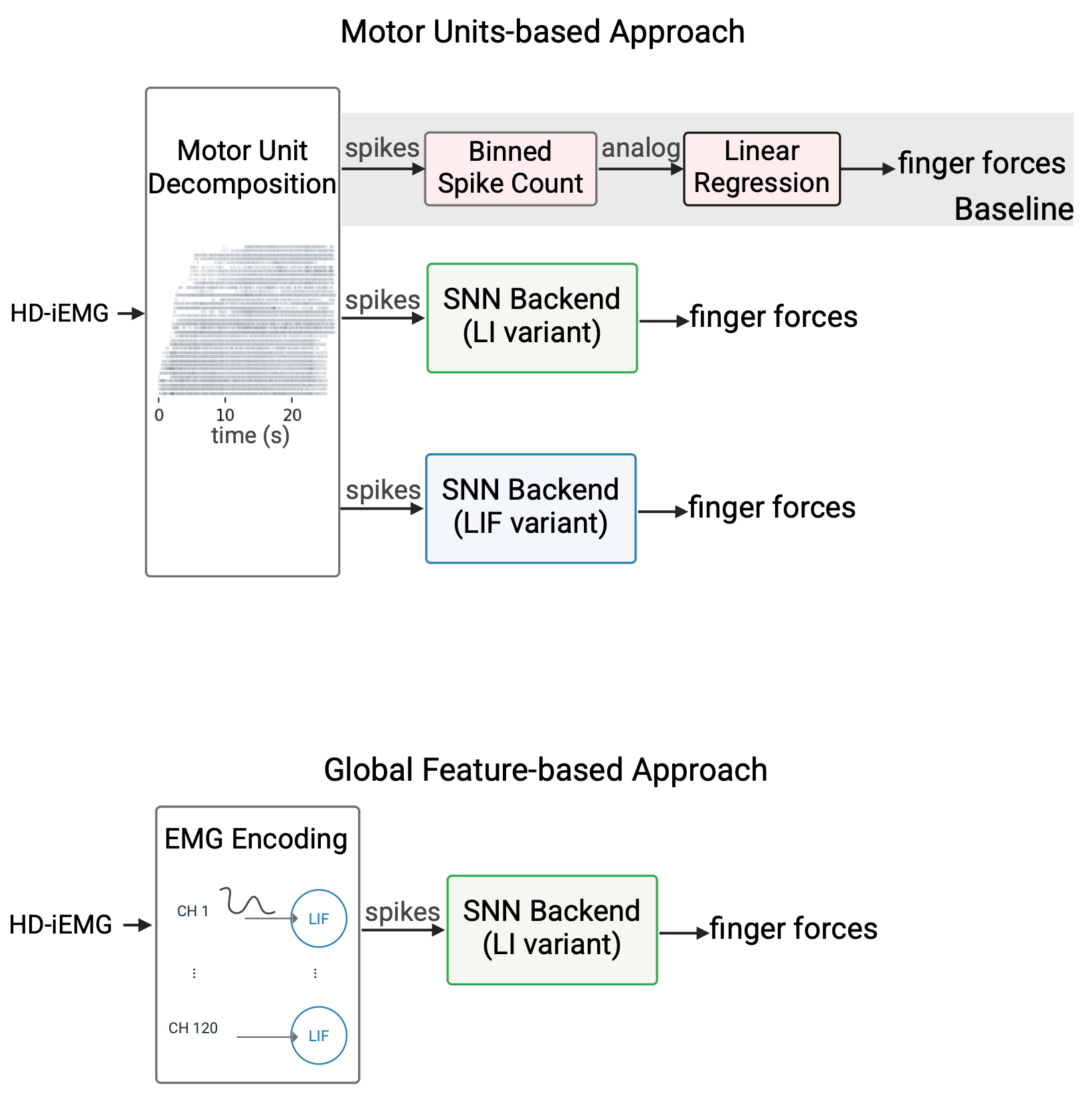}
            \caption{}
            \label{fig:experiments_overview}
        \end{subfigure}
        \end{adjustbox}
    \end{minipage}
    \hfill
    \begin{minipage}[b]{0.57\textwidth}
        \begin{subfigure}[c]{\linewidth}
            \centering
            \includegraphics[width=\linewidth]{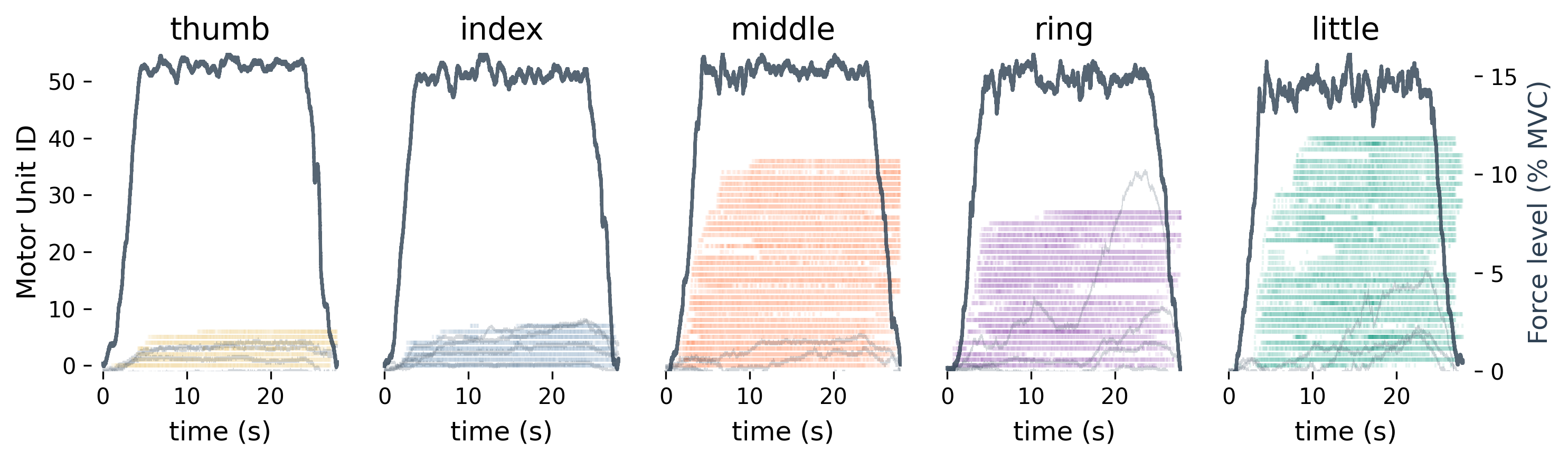}
            \caption{}
            \label{fig:example_decomposition}
        \end{subfigure}
        \begin{subfigure}[b]{\linewidth}
            \centering
            \includegraphics[width=\linewidth]{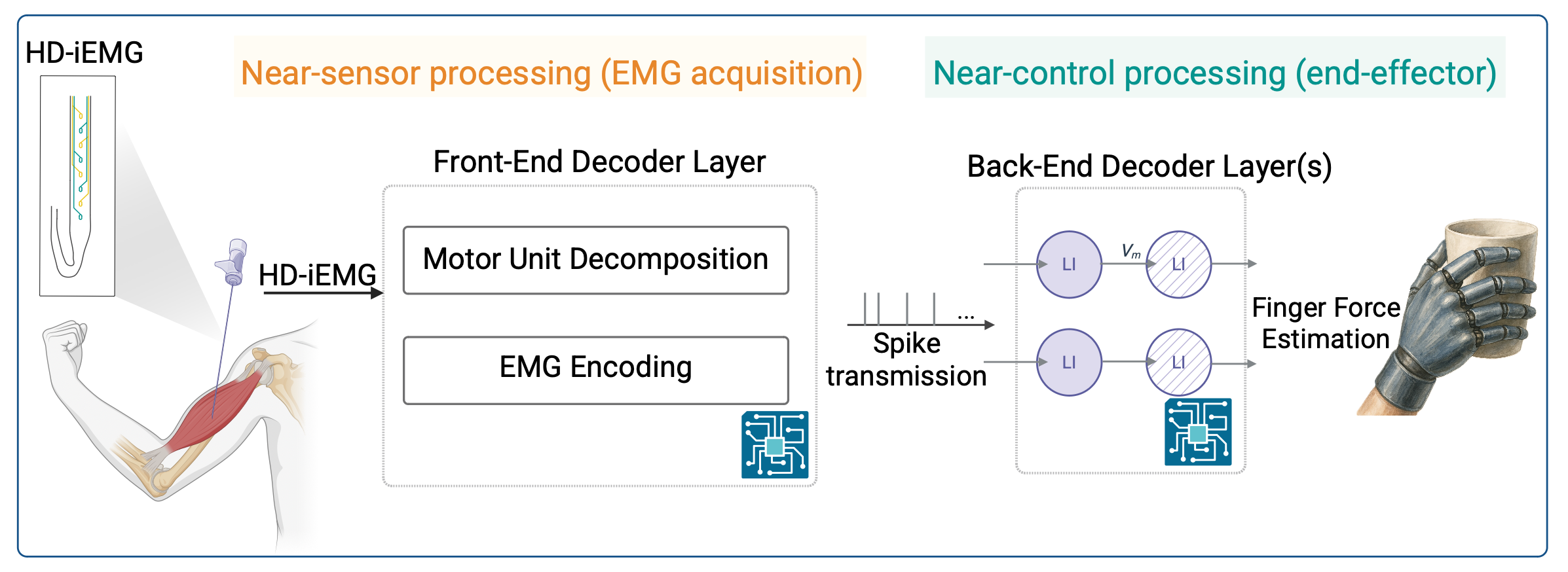}
            \caption{}
            \label{fig:envisioned_pipeline}
        \end{subfigure}
    \end{minipage}
\caption{\textbf{Force decoding from \ac{HD-iEMG} using spike-based representations.} (a) Schematic of decoding architectures using either decomposed motor unit spike trains (top) or spike-encoded \ac{HD-iEMG} signals (bottom). Motor unit spikes are processed through a \ac{LI} layer or a cascade of \ac{LIF} neurons followed by \ac{LI} units. The baseline model applies linear regression to binned spike counts. (b) Example task-specific motor unit activity decomposed from \ac{HD-iEMG} and corresponding force output during isometric contractions across five finger tasks (Subject~1, repetition~2). Motor units were decomposed independently for each task; therefore, motor unit IDs (y-axis) are not shared across the five finger tasks. (c) Conceptual overview of the proposed pipeline implementation for near-sensor spike-based processing and distal force estimation.}
\label{fig:general_framework}
\end{figure}
We demonstrate the potential of spike-based processing for decoding individual finger movements in isometric contraction at 15\% \ac{MVC}, exploiting a novel \ac{HD-iEMG} setup~\cite{grison_intramuscular_2024}. We constructed two \ac{SNN} architectures and investigated two possible input modalities derived from intramuscular \ac{EMG} recordings: decomposed motor unit spike trains and direct spike-encoded representations (Fig.~\ref{fig:experiments_overview}). For both cases, the resulting input spikes are streamed into the \acp{SNN}  composed of \ac{LIF} neurons and biophysically realistic synapses which leverage their intrinsic temporal dynamics to process continuous force trajectories, without requiring extra spike binning or stand-alone feature extraction stages. 

We investigated two variants of the pipelines, distinguished by the readout neuron type of the decoder network. In the leaky readout variant, motor units were mapped to five \ac{LI} units through dense feedforward synapses, with continuous force estimates for each finger read directly from their continuous time real-value output. In the spiking variant, a first layer of five \ac{LIF} neurons processed the motor unit spike trains, followed by a cascade of two \ac{LI} layers connected one-to-one with the preceding units: the first transformed spike trains into continuous values, and the second applied temporal smoothing to yield the final force estimates. Hereafter, we refer to these two configurations as the \ac{LI} and \ac{LIF} configurations, respectively.

We conducted three experiments to: (1) benchmark the two \ac{SNN} topologies against a conventional linear baseline for decoding finger forces from motor unit spike trains; (2) assess model robustness to spike omissions simulating decomposition errors; and (3) compare decoding performance using spike-encoded \ac{HD-iEMG} signals as an alternative to motor unit-based inputs.

\subsection{Streamed processing of motor units spike trains with neurons and synapses} \label{ssec:res_spike_based_processing}
Three high-density intramuscular \acp{MEA} with a total of 120 channels were inserted within the forearm muscles of two healthy male participants as they performed sequential isometric flexion and extension tasks with individual fingers at 15\% of their \ac{MVC}. Motor unit spike trains were decoded from the recorded intramuscular \ac{HD-iEMG} using a validated blind source separation algorithm designed for high-yield decomposition~\cite{grison_intramuscular_2024, grison_particle_2024, grison_unlocking_2025}. Decomposition was performed independently for each finger and movement direction, and the resulting spike trains were aggregated to form direction-specific input sets. For instance, all motor units identified during flexion trials of all fingers were combined to train the decoder. A total of 121 motor units were identified during finger flexion tasks and 196 during extension tasks for Subject~1. For Subject~2, 51 units were identified during flexion and 93 units during extension. An example of decomposed motor units spike trains during finger extension tasks is shown in Fig.~\ref{fig:example_decomposition}.

\begin{figure}[ht!]
    \centering
    \begin{minipage}[b][\height][c]{0.62\linewidth}
        \begin{subfigure}{\linewidth}
            \centering
            \includegraphics[width=\linewidth]{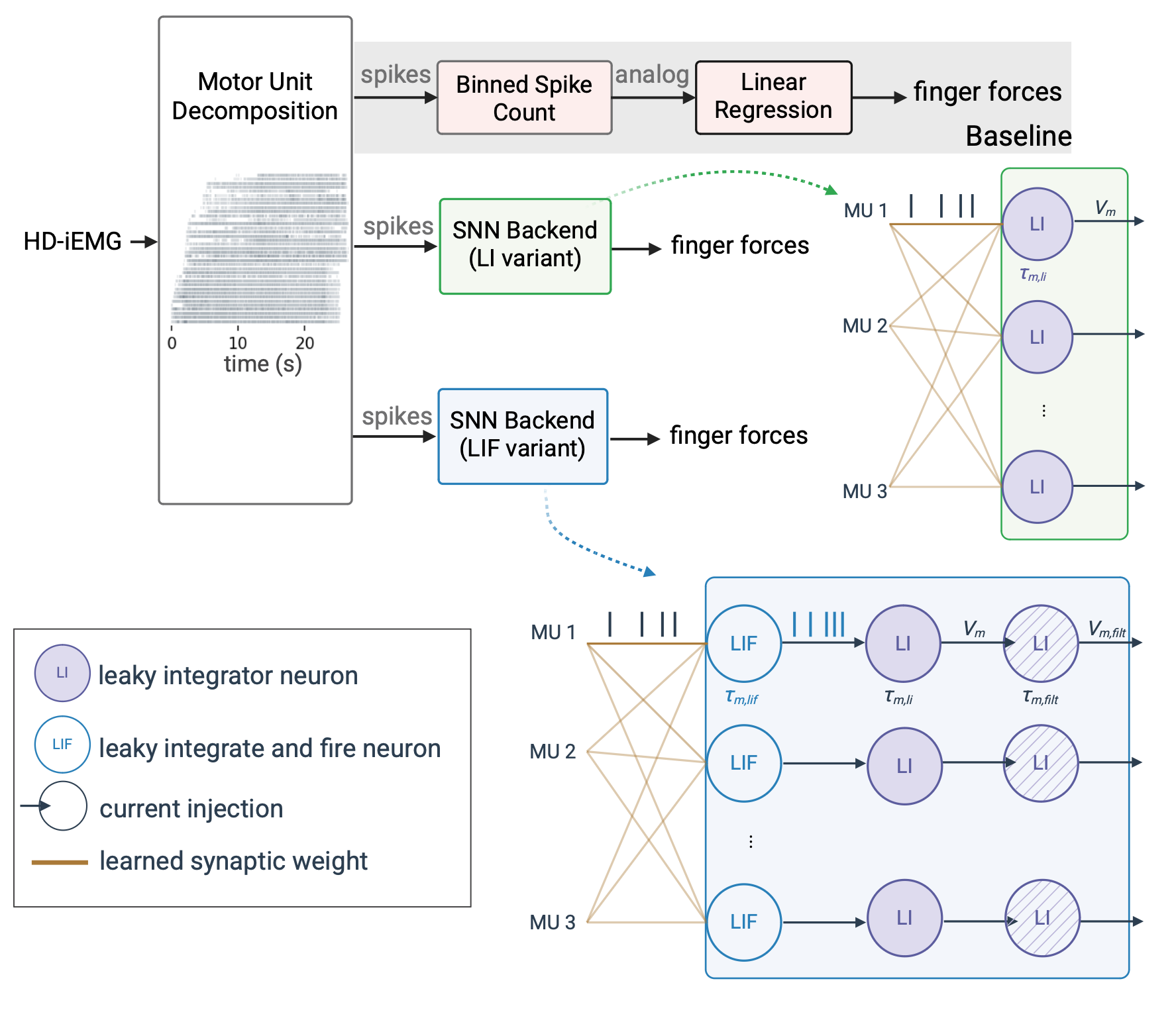}
            \caption{}
            \label{fig:mu_versus_count_schematic}
        \end{subfigure}
    \end{minipage}
    \hfill
        \begin{minipage}[b]{0.33\linewidth}
        \begin{subfigure}[b]{\linewidth}
            \includegraphics[width=\linewidth]{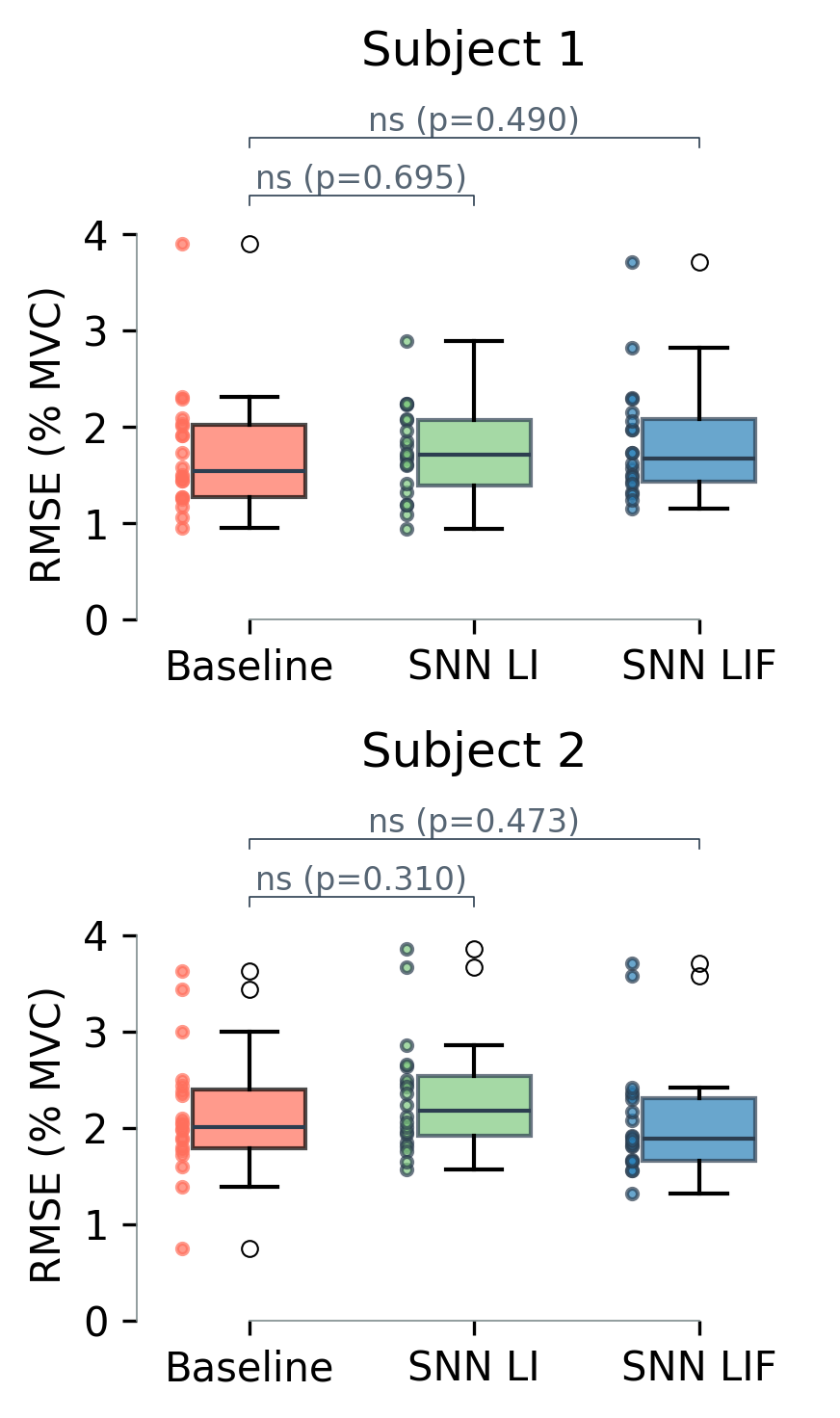}
            \caption{}
            \label{fig:boxplot_comparison}
        \end{subfigure}
    \end{minipage}

    \caption{\textbf{Comparison of motor unit-based decoder performance across a baseline linear regression model and \ac{SNN}} (a) Schematic illustration of the two decoding pipelines: motor unit spike counts binned over fixed windows and decoded with a linear regression model (baseline), versus direct decoding from motor unit spike times with \ac{SNN} backends. (b) \ac{RMSE} distributions for individual finger flexion and extension tasks, comparing the baseline model (spike count) to SNN decoders with leaky integrator (\ac{LI}) and leaky integrate-and-fire (\ac{LIF}) readout units. Points represent \ac{RMSE} values for each repetition, finger task, and movement direction; p-values from two-sided Mann–Whitney U tests are reported for each pairwise comparison.}
    \label{fig:snn_to_lr_15_comparison}
\end{figure}

In the first experiment, we compared \ac{SNN} decoders to a conventional linear regression baseline, as used in previous studies~\cite{xu_estimation_2021, xia_simultaneous_2025}. The overview schematic for the first experiment is illustrated in Fig.~\ref{fig:mu_versus_count_schematic}. The baseline model received binned spike counts computed over \SI{80}{\milli\second} analysis windows with 50\% overlap as input features (see Methods~\ref{ssec:mbaseline}). In contrast, the \ac{SNN} decoders operated directly on the motor unit spike trains, which were streamed into a dense layer of either spiking neurons or leaky integrator units. This first layer integrated input spikes continuously, as they arrived, without requiring explicit temporal segmentation (Fig.~\ref{fig:envisioned_pipeline}).

All models were trained in a multi-output regression setting, simultaneously estimating the force trajectories of all five fingers from the pool of identified motor units associated with a given movement direction (flexion or extension). Performance metrics, summarized in Fig.~\ref{fig:boxplot_comparison}, report \ac{RMSE} values averaged across two repetitions using cross-validation and computed over all finger tasks. Additional metrics, including \ac{R2} and \ac{MAE}, are summarized in Table~\ref{tab:summary_decoder_performance} (see Supplementary Table~\ref{tabsupp:mu_directional_performance_full} for a detailed performance breakdown by movement direction). The baseline linear regression model achieved an average \ac{RMSE} of $1.73~\pm~0.29$ \%\ac{MVC} for Subject~1 and $2.13~\pm~0.32$ \%\ac{MVC} for Subject~2. The \ac{LI} configuration achieved $1.73~\pm~0.23$ and $2.31~\pm~0.34$ \%\ac{MVC}, while the \ac{LIF} configuration yielded $1.78~\pm~0.27$ and $2.12~\pm~0.41$ \%\ac{MVC} for Subjects~1 and~2, respectively. Both network variants achieved decoding performance comparable to the baseline model.

\begin{table}[h!]
\centering
\caption{Decoder performance across subjects, input feature and models (\ac{LR} and \acp{SNN}). Metrics are averaged across fingers, directions, and 10 random initialization seeds for the \ac{SNN} decoders.}
\begin{tabular}{c|c|p{2.8cm}|c|c|c}
\toprule
\textbf{Subject} & \textbf{Input Type} & \textbf{Decoder Model} & \textbf{RMSE} (\%\ac{MVC}) & \textbf{MAE} (\%\ac{MVC}) & \textbf{R$^2$} \\
\midrule
\multirow{3}{*}{1}
& \textbf{motor unit spike count} & \textbf{Baseline (LR)} & $\mathbf{1.73~\pm~0.29}$ & $\mathbf{1.01~\pm~0.14}$ & $\mathbf{0.88~\pm~0.04}$ \\ \cmidrule{2-6}
& \multirow{2}{*}{motor unit spike train} & \ac{SNN} (\ac{LI}) & $1.73~\pm~0.23$ & $1.09~\pm~0.13$ & $0.89~\pm~0.03$ \\
&  & \ac{SNN} (\ac{LIF}) & $1.78~\pm~0.27$ & $1.10~\pm~0.21$ & $0.88~\pm~0.04$ \\
\cmidrule{2-6}
& encoded \ac{HD-iEMG} & \ac{SNN} (\ac{LI}) & $2.55~\pm~1.13$ & $1.72~\pm~0.75$ & $0.71~\pm~0.22$ \\
\midrule
\multirow{3}{*}{2}
& \textbf{motor unit spike count}& \textbf{Baseline (LR)} & $\mathbf{2.13~\pm~0.32}$ & $\mathbf{1.36~\pm~0.16}$ & $\mathbf{0.84~\pm~0.05}$ \\ \cmidrule{2-6}
& \multirow{2}{*}{motor unit spike train}  & \ac{SNN} (\ac{LI}) & $2.31~\pm~0.34$ & $1.64~\pm~0.29$ & $0.82~\pm~0.05$ \\
&  & \ac{SNN} (\ac{LIF}) & $2.12~\pm~0.41$ & $1.46~\pm~0.38$ & $0.85~\pm~0.06$ \\
\cmidrule{2-6}
& encoded \ac{HD-iEMG} & \ac{SNN} (\ac{LI}) & $3.41~\pm~1.23$ & $2.38~\pm~0.67$ & $0.57~\pm~0.32$ \\
\bottomrule
\end{tabular}
\label{tab:summary_decoder_performance}
\end{table}

\begin{figure}
    \centering
    \includegraphics[width=1\linewidth]{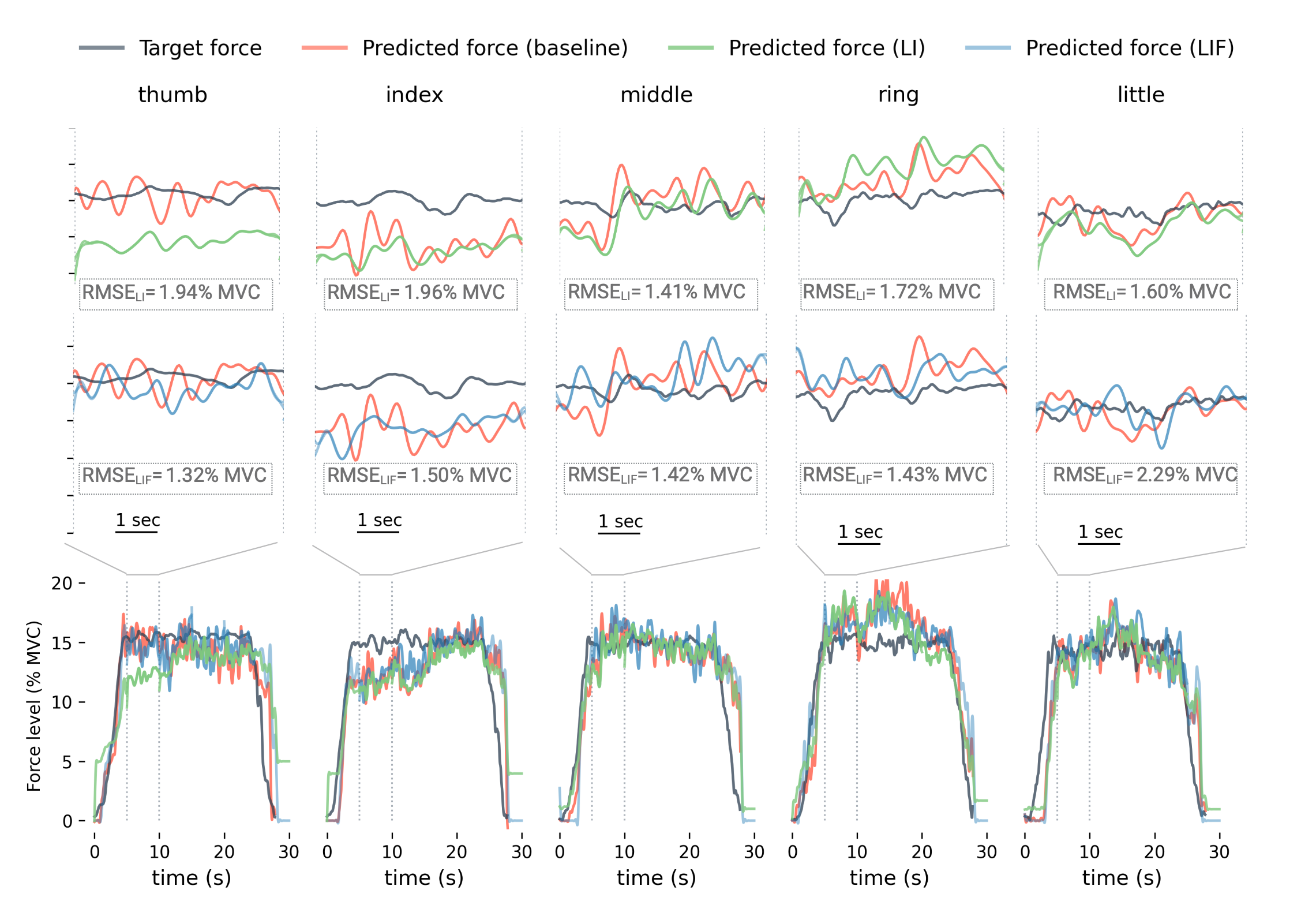}
    \caption{Example decoding performance of the baseline and \ac{SNN} models during finger flexion tasks. Representative force estimation traces for Subject~1 (repetition~2) using motor unit spike trains as input. The five columns correspond to individual finger flexion tasks (thumb to little finger). Top insets: zoomed view of the 5–10 s interval, showing target force (black) alongside predicted forces from the baseline linear regression model (red), the \ac{LI} configuration (green), and the \ac{LIF} configuration (blue). \ac{RMSE} (\%\ac{MVC}) for each model is reported in each inset. Bottom panels: full 30 s contraction profiles for each task, with traces for non-targeted fingers omitted for clarity.}
    \label{fig:predictions_comparison}
\end{figure}

Representative decoding outputs from single trials of the five finger tasks are shown in Fig.~\ref{fig:predictions_comparison}. Each column corresponds to a different finger, and for clarity, only the trajectory of the actively engaged (i.e., highly activated) finger is depicted. Although all models decoded forces for all five fingers simultaneously, non-target finger outputs are omitted for visual clarity. Across all models, the overall shape of the force profile—defined by the rising, plateau and falling phases of the isometric contractions—was reliably captured. However, finer modulations in force during the plateau phase were not consistently reproduced, as highlighted in the zoomed five-second inset. This limitation likely stems from an incomplete sampling of the motor unit activity required to recover those fluctuations. As a consequence, subtle variations in motor unit discharge rates encoding fine force adjustments may have been partially lost~\cite{keenan_influence_2005}, thereby limiting the decoder’s ability to track slow force changes during sustained contractions. While this limitation may pose challenges for applications that demand high-fidelity force reconstruction—such as precision control in virtual reality scenarios—it is likely less critical in assistive contexts like prosthetic hand control for amputees, where capturing fine fluctuations is often unnecessary. Nevertheless, further analysis is needed to assess the functional implications of this limitation and to explore strategies for improving resolution during sustained-force decoding from motor units.

We also acknowledge that the limited sample size constrains the statistical power of these comparisons. Nonetheless, the results suggest that shallow \ac{SNN} architectures composed of \ac{LI} units or \ac{LIF} neurons can in principle achieve decoding performance on par with conventional regression pipelines using the decomposed motor unit spike trains. A key advantage of the approach presented lies in its event-driven, continuous processing capability, which eliminates the need for explicit memory buffers or spike count accumulation. Instead, the network leverages the intrinsic state dynamics of the neurons, which integrate and decay over time according to their membrane time constants.

\begin{figure}[ht]
    \begin{subfigure}[]{\linewidth}
         {\includegraphics[width=1\linewidth]{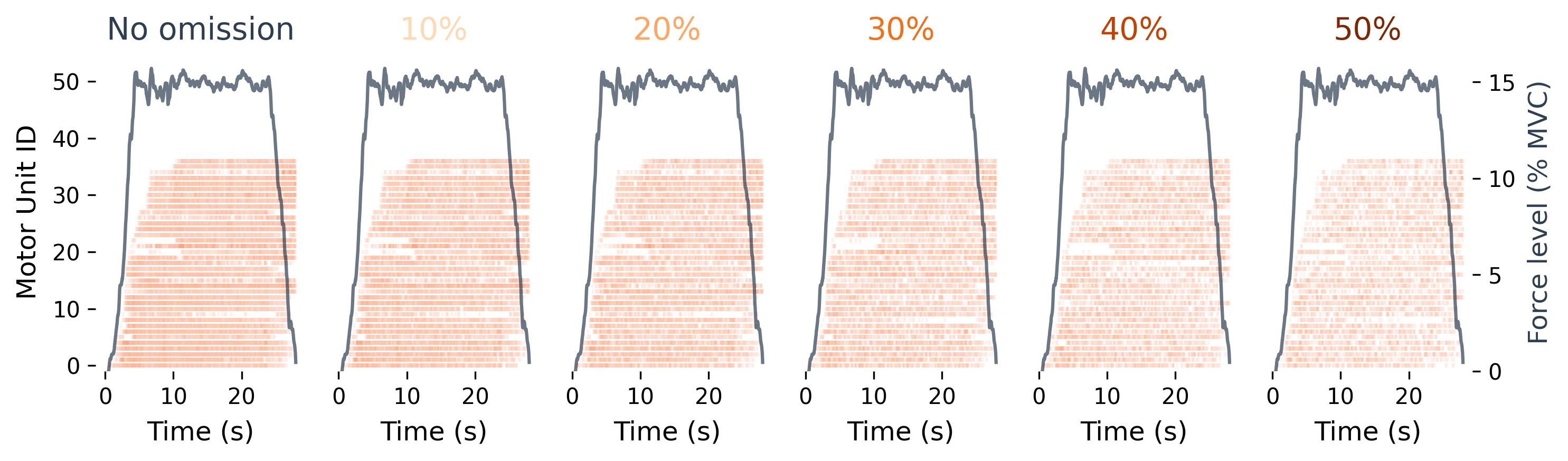}}
         \caption{}
         \label{fig:raster_across_omissions}
    \end{subfigure}

    \begin{subfigure}[]{\linewidth}
        {\includegraphics[width=1\linewidth]{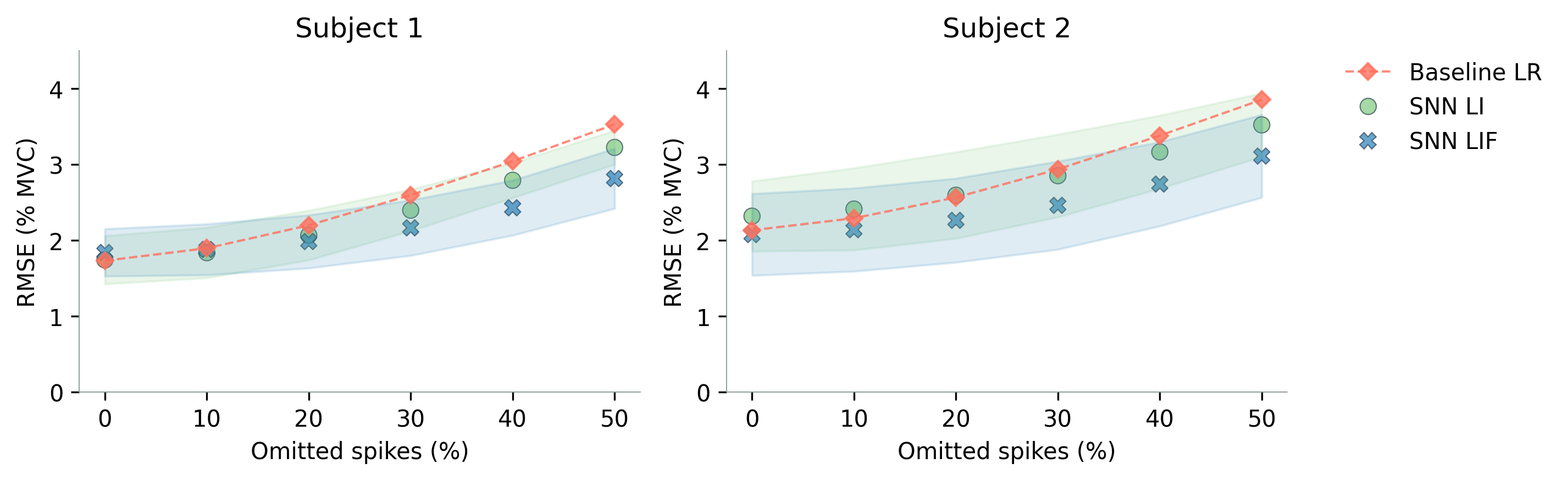}}
        \caption{}
        \label{fig:metrics_across_omissions}
    \end{subfigure}
    \caption{\textbf{Robustness of \ac{SNN} configurations to random spike omission during inference.} (a) Raster plots of decomposed motor unit spike trains from a representative trial (middle finger flexion, repetition~2, Subject~1) under increasing levels of motor units spike omission (from 0\% to 50\%). Spike omissions were applied randomly during inference to simulate decomposition noise. Overlaid force trajectories (\%\ac{MVC}) correspond to the target force level during the task.
    (b) \ac{RMSE} decoding performance across all tasks and movement directions for both subjects as a function of spike omission level. Models were trained on clean motor unit spike trains (i.e. no omission) and evaluated under varying levels of omission. Shaded areas indicate standard deviation across finger tasks and directions. The standard deviation of the baseline model were omitted for clarity and are reported in Supplementary Table~\ref{tabsupp:spike_omission_robustness}.}
    \label{fig:spike_omission}
\end{figure}

\subsection{Networks resilience to real-time decomposition errors}\label{ssec:res_robustness}
Real-time decomposition of motor unit activity typically relies on precomputed separation matrices, derived offline, which are subsequently applied to streamed \ac{EMG} signals for online motor unit identification~\cite{barsakcioglu_control_2021, rossato_i-spin_2024}. This strategy circumvents the computational cost of performing blind source separation in real time, but introduces trade-offs: variations in motor behavior, electrode placement, or signal quality between calibration phase and online phase can influence both the number of motor units detected and the temporal precision of their identified discharge timings.

To simulate the effects of these limitations in a controlled setting, we introduced perturbations to the input motor unit spike trains in the form of random spike omissions. This approach emulates a subset of identification errors likely encountered during real-time operation-specifically, false negatives, in which motor unit spikes are missed~\cite{kline_error_2014}. Rather than explicitly retraining or recalibrating the decoders, we evaluated the impact of these perturbations at inference time, to quantitatively assess the intrinsic noise resilience of the baseline model and \ac{SNN} decoding architectures.

We hypothesized that the spiking intermediate layer configuration would confer greater robustness to these omissions compared to both the baseline model and the leaky direct-readout configuration, due to the thresholding properties of \ac{LIF} neurons. By filtering out small membrane fluctuations induced by missing spikes, spiking neurons can inherently stabilize the network output against noisy inputs.

Figure~\ref{fig:spike_omission} illustrates the degradation in decoding performance, quantified by RMSE, under increasing levels of spike omissions (see Supplementary Table~\ref{tabsupp:spike_omission_robustness} for corresponding MAE and R$^2$ metrics). As expected, performance declined with higher omission rates across all models; however, the \ac{LIF} configuration consistently exhibited greater robustness compared to both the baseline linear model and the \ac{LI} configuration. In Subject~1, for example, the RMSE of the baseline model increased from $1.73~\pm~0.26$ to $3.53~\pm~0.29$~\%\ac{MVC} at 50\% spike omission, while the \ac{LI} model rose from $1.74 \pm 0.32$ to $3.22~\pm~0.22$~\%\ac{MVC}. The \ac{LIF} model showed a more moderate increase from $1.84~\pm~0.31$ to $2.81~\pm~0.40$~\%\ac{MVC}, indicating improved resilience to spike loss.

This robustness, however, comes at the cost of a slight increase in parameter space and, consequently, memory footprint. As shown in Fig.~\ref{fig:tradeoff_rmse_footprint} and detailed in Supplementary Table~\ref{tabsupp:static_metrics}, the memory required to store model parameters is marginally higher for the \ac{LIF} decoder compared to the \ac{LI} configuration. In Subject~1, for example, the parameter memory footprint was 98,036 bytes for the \ac{LI} configuration and 98,092 bytes for \ac{LIF}. It is important to note that these values represent a conservative, worst-case estimate, as they assume full 32-bit floating-point precision for all parameters. In practice, memory requirements can be substantially reduced through quantization, albeit at the potential cost of reduced performance. However, part of this performance loss can be mitigated by bio-inspired mechanisms, such as introducing heterogeneity and population coding~\cite{baracat_finger_2025}. A systematic analysis of the trade-off between memory optimization and decoding accuracy, however, lies beyond the scope of this study and is left for future work.

\begin{figure}
    \centering
    \includegraphics[width=0.8\linewidth]{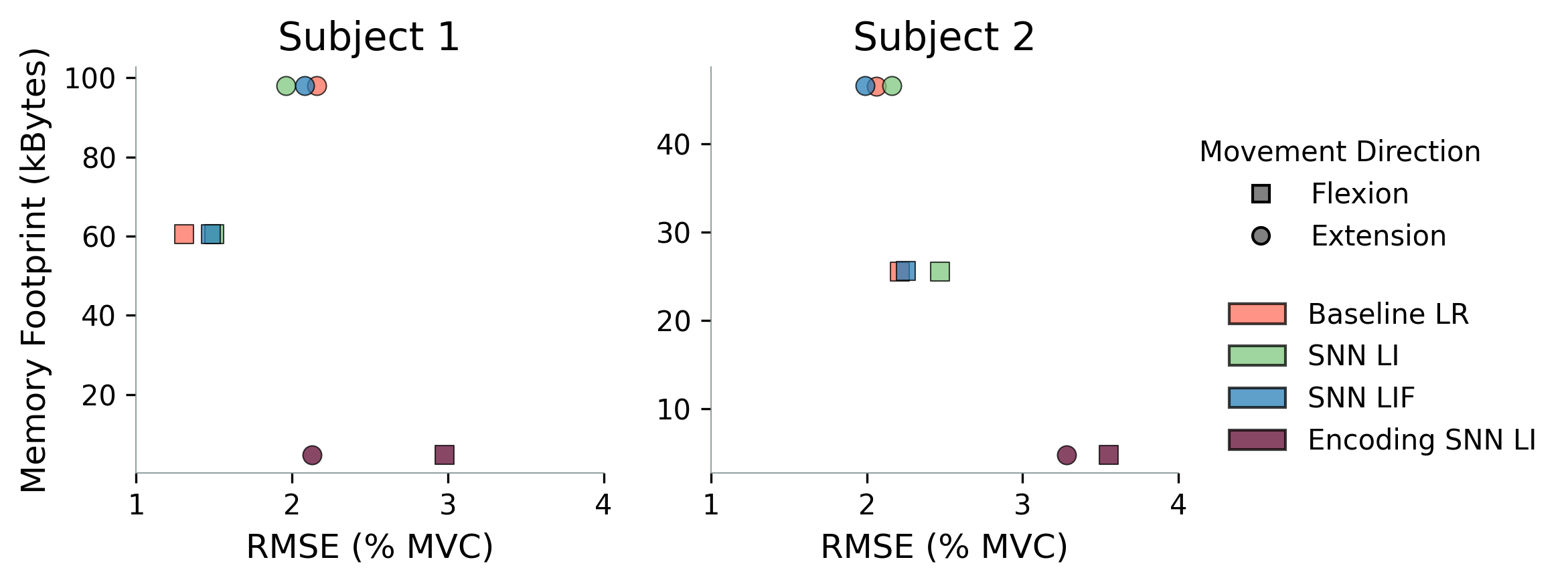}
    \caption{Relation between decoder accuracy, measured as \ac{RMSE}, expressed as \%\ac{MVC}), and memory footprint (in kilobytes). Memory estimates account for linear coefficients in the baseline model and, for \acp{SNN}, include synaptic weights, biases, and time constants. For decoders based on motor unit spike trains, the memory footprint also includes the separation matrices required for decomposition.}
    \label{fig:tradeoff_rmse_footprint}
\end{figure}

\subsection{Encoding HD-iEMG into events bypassing motor unit decomposition}\label{ssec:res_encoding}
\begin{figure}[ht!]
     \begin{subfigure}[b]{0.7\linewidth}
        {\includegraphics[width=\linewidth]{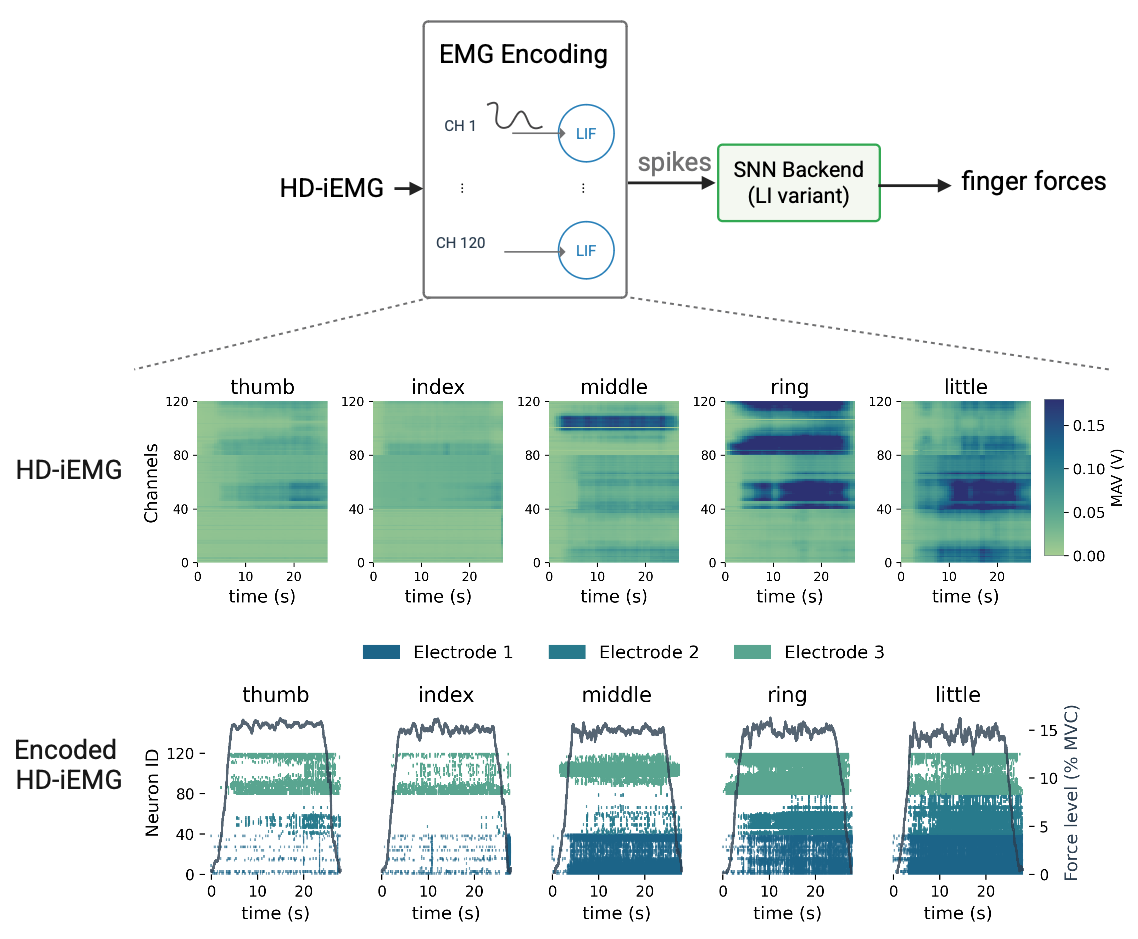}}
        \caption{}
        \label{fig:example_encoding_flexion}
    \end{subfigure}
    \hfill
    \begin{subfigure}[b]{0.29\linewidth}
    \includegraphics[width=\linewidth]{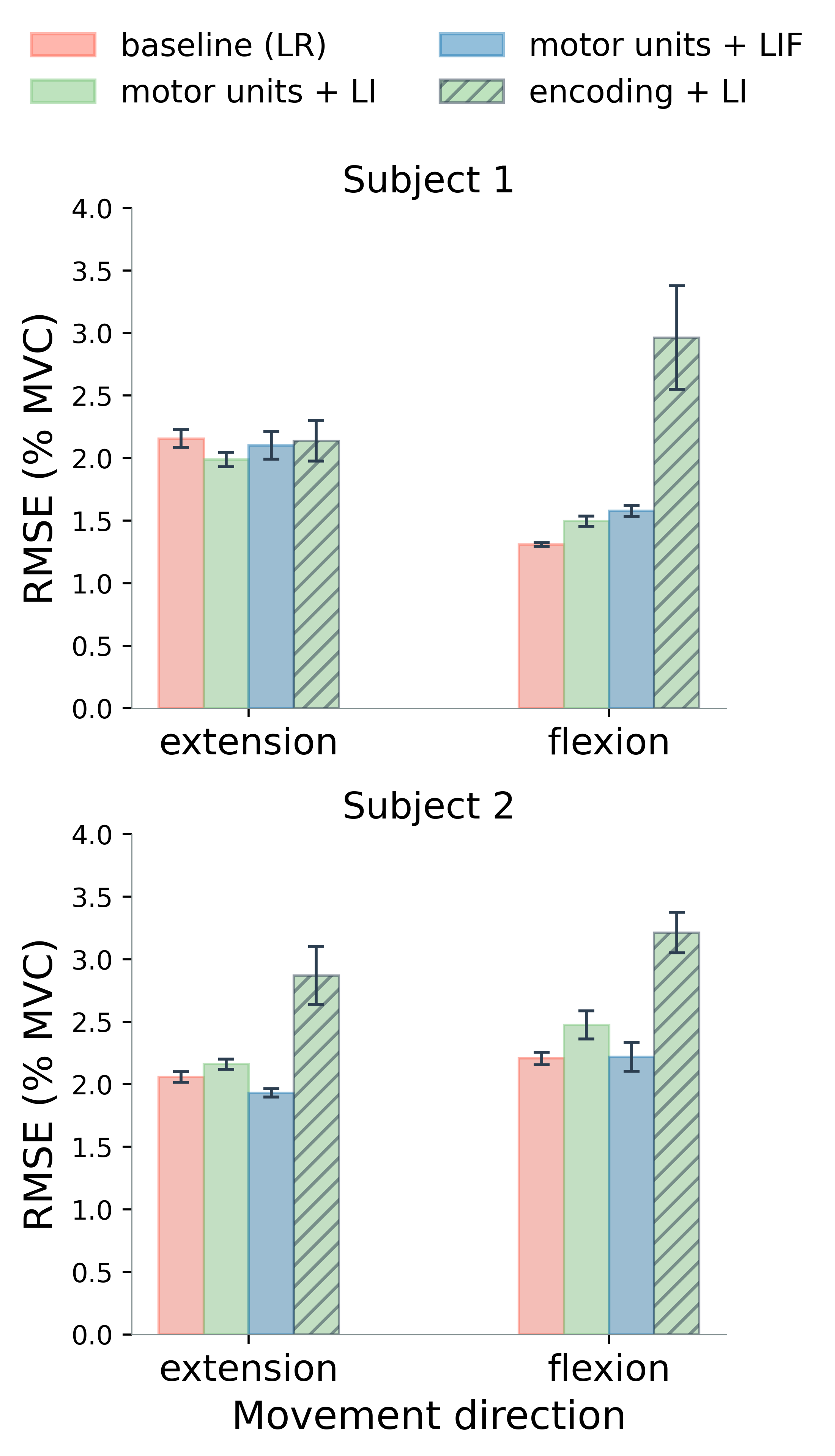}
    \caption{}
    \label{fig:encoding_to_mu_comparison}
    \end{subfigure}
\caption{\textbf{Decoder performance using spike-encoded \ac{HD-iEMG} signals.} 
(a) (Top) Schematic of the processing pipeline for decoding finger forces from \ac{HD-iEMG}: filtered signals from 120 channels are spike-encoded using \ac{LIF} neurons and passed to an \ac{SNN} backend. (Middle) Input \ac{iEMG} signals for each finger task smoothened with a \ac{MAV} on windows of 1-second and 50\% overlap. Each column represents one finger task, and all channels from the three electrodes (120 total) are included.
(Bottom) Example raster plots of spikes generated by the encoding \ac{LIF} neurons in response to the \ac{iEMG} inputs (Subject~1, flexion tasks, repetition~2). Force trajectories (in \%\ac{MVC}) are overlaid for reference. Neurons are color-coded by electrode. Spike encoding was performed using \ac{LIF} neurons with a homogeneous membrane time constant of \SI{20}{\milli\second}, while thresholds were adjusted per electrode to account for amplitude differences (Electrode 1: \SI{0.1}{\volt}, Electrode 2: \SI{0.4}{\volt}, Electrode 3: \SI{0.2}{\volt}). 
(b) Decoding performance reported as mean \ac{RMSE} across finger tasks for each subject, separated by movement direction.}
\label{fig:encoding_to_mu_results}
\end{figure}

In parallel with the use of decomposed motor unit spike trains, we investigated in a third experiment whether direct event-based encoding of \ac{HD-iEMG} signals could support proportional and simultaneous decoding of finger forces (Fig.~\ref{fig:example_encoding_flexion}, top panel). This approach overcomes the computational overhead associated with offline motor unit decomposition and eliminates the need for storing separation matrices during real-time operation, offering a more lightweight and practical alternative for wearable implementations. However, because this strategy essentially relies on global features extracted from \ac{EMG} activity rather than direct neural discharge information, a reduction in decoding accuracy relative to motor unit-based approaches was anticipated. The objective of this experiment is to quantify the extent of information loss incurred when bypassing decomposition, and to understand whether the trade-off between performance and implementation feasibility could justify favoring this strategy for real-time control scenarios.

Filtered \ac{HD-iEMG} signals were encoded into spike trains using an additional layer of \ac{LIF} neurons configured with homogeneous membrane time constants and electrode-specific spiking thresholds. Thresholds were individually tuned per electrode for Subject~1 to account for the greater variability in signal amplitude arising from differences in electrode placement across muscles, but kept homogeneous in Subject~2. Although the resulting spike trains do not represent physiological neural discharge patterns, they provide a binary, event-based representation of muscle activity suitable for neuromorphic processing and \ac{SNN} backend. An example of the spike-encoded representation during finger flexion tasks is shown in Fig.~\ref{fig:example_encoding_flexion} (bottom panel), alongside the corresponding \acp{MAV} features of the original filtered \ac{iEMG} signals (Fig.~\ref{fig:example_encoding_flexion}, middle panel). A similar spatial activation pattern is observed between the two representations. Notably, the electrode-specific threshold tuning introduced during encoding successfully normalized the spike activity across channels, mitigating the inter-electrode amplitude variability still visible in the original \ac{MAV} feature map. Channels corresponding to the middle electrode (channels 40–80) showed higher activation during finger extension tasks, consistent with their anatomical targeting. Accordingly, higher thresholds were assigned to these channels to balance the spike activity across the array.

\begin{figure}[ht!]
 \includegraphics[width=1\linewidth]{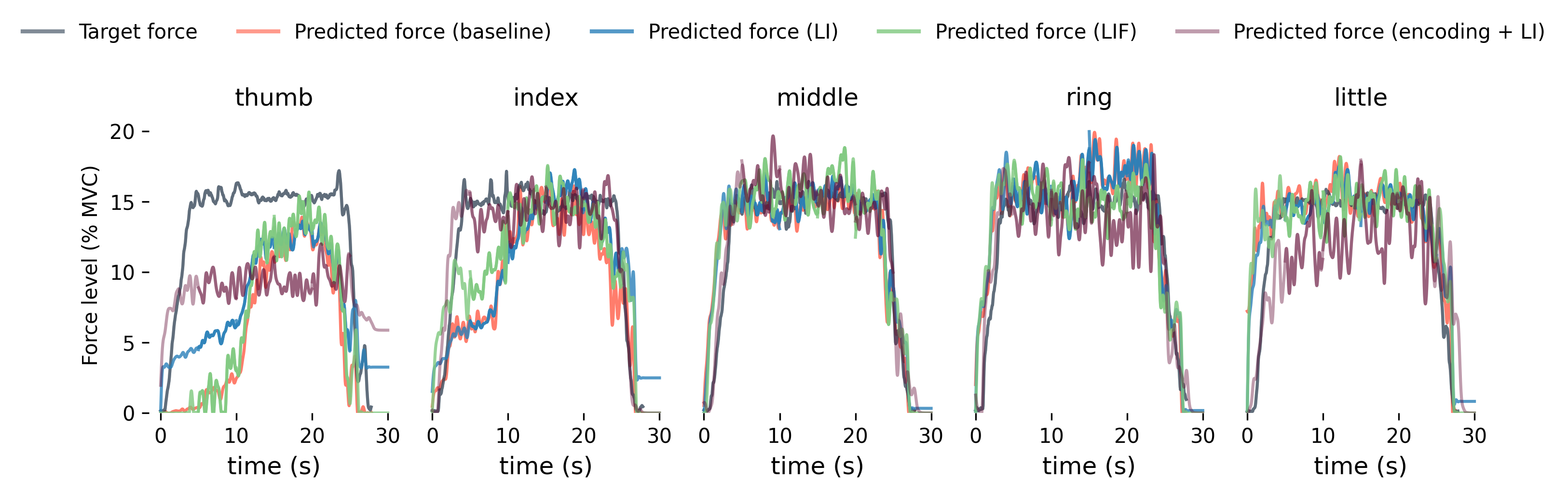}
\caption{Representative decoding of finger extension forces from spike-based inputs.
Force estimation traces for Subject~1 (repetition~1) across decoder configurations. The \ac{LI} (blue) and \ac{LIF} (green) models received decomposed motor unit spike trains; the encoding + \ac{LI} model (purple) used spike-encoded \ac{HD-iEMG}; and the baseline linear regression decoder used binned spike counts of motor units as input.}
\label{fig:example_encoding_prediction}
\end{figure}

After encoding, the resulting spike trains were directly streamed into the same leaky \ac{SNN} architecture used for motor unit-based decoding, for a systematic comparison between the two input modalities under identical network configuration. Importantly, the number of trainable parameters remained unchanged; the encoding layer simply replaced physiological motor unit spike timing information with a binary threshold-based representation, without increasing model complexity, defined in terms of the number of trainable parameters.

Performance metrics for motor unit-based decoding were previously reported (see Section~\ref{ssec:res_spike_based_processing}). In comparison, decoding based on spike-encoded \ac{HD-iEMG} signals resulted in higher \ac{RMSE} values as expected, with the leaky configuration reaching $2.55~\pm~1.13$ \%\ac{MVC} for Subject~1 and $3.41~\pm~1.23$ \%\ac{MVC} for Subject~2 (see Table~\ref{tab:summary_decoder_performance}). This represents a performance drop of approximately 0.8–1.1~\%\ac{MVC} relative to motor unit-based \ac{SNN} decoders. Interestingly, most of the error increase originated from the flexion tasks, as shown in Fig.~\ref{fig:encoding_to_mu_comparison}. One likely explanation is the suboptimal tuning of the encoding layer parameters. For practical reasons, we kept these parameters fixed across both movement directions to mimic real-time constraints. However, the underlying \ac{HD-iEMG} amplitude distributions differ substantially between flexion and extension depending on the electrodes' positions~\cite{grison_intramuscular_2024}, which have resulted in different event rates across movement directions (see Supplementary Table ~\ref{tabsupp:mu_decomposition}). Signal amplitude variability was also observed across finger tasks within the same movement direction, with some tasks producing sparse event streams while others yield dense ones (see Supplementary Fig.~\ref{figsupp:encoding_input_emg_flexion}). This variability in the encoded representations directly impacted decoding accuracy, as illustrated in Fig.~\ref{fig:example_encoding_prediction}, where the thumb and index tasks exhibited the lowest prediction accuracy—corresponding to reduced spike activity in the encoded input.

Despite the drop in accuracy, spike-encoded \ac{HD-iEMG} representations offer a compelling tradeoff in terms of memory efficiency. Unlike motor unit-based decoders, they do not require storage of decomposition separation matrices, which scale with the number of extracted motor units and the number of \ac{EMG} channels. This memory overhead can be substantial and varies significantly between subjects depending on the number of decomposed motor units, whereas the encoding approach ensures a fixed memory allocation. For example, in Subject~1, the flexion decoder required an additional 58,080 bytes, assuming full-resolution (32-bit float) representation, solely to store the separation matrix (based on 121 identified units), as shown in Fig.~\ref{fig:tradeoff_rmse_footprint} and detailed in Supplementary Table~\ref{tabsupp:static_metrics}.

Together, these observations highlight a fundamental tradeoff: while spike-encoded \ac{HD-iEMG} inputs result in slightly reduced accuracy, they achieve that with lower memory demands. Importantly, the gap in performance is not inherent and could be reduced through improved encoding strategies. Future implementations may benefit from adaptive thresholding~\cite{Sharifshazileh_Indiveri23} or heterogeneous encoding populations~\cite{baracat_heterogeneous_2025} to better capture the input signal variability. Ultimately, the choice between physiologically plausible motor unit inputs and efficient spike-based encoding should be guided by the specific accuracy, memory requirements of the intended application.

\section{Discussion}\label{sec:discusion}
Proportional and simultaneous decoding of individual finger forces from \ac{EMG} signals is a key milestone towards naturalistic motor control for assistive technologies. Attaining this level of fine control requires decoding systems that are sufficiently accurate, but also ones that can operate with low-latency and high-energy efficiency. In this study, we combined and integrated the latest state-of-the-art results in \ac{HD-iEMG} signal acquisition (high-density intramuscular \acp{MEA}), motor unit decomposition algorithms, and \ac{SNN} to build a pipeline able to simultaneously decode individual finger force trajectories, and to demonstrate the potential of \acp{SNN} in this domain. We systematically and quantitatively assessed the impact of key architectural decisions on performance accuracy, noise robustness, and memory footprint. Our experiments focused on \ac{HD-iEMG} to ensure high-quality decomposition; however, the overall framework and comparative analyses remain, in principle, applicable to \ac{HD-sEMG} recordings.

Neuromorphic computing has recently gained traction for efficient implementations of \ac{iBMI} neural activity decoding systems~\cite{biyan_combining_2025, Gallou_etal24, shaikh_real-time_2019}. These developments offer a promising solution to two emerging challenges. First, achieving finer-grained decoding of user intent requires denser electrode arrays, resulting in a dramatic increase in data bandwidth. However, transmitting this high-volume data wirelessly from implants remains impractical due to stringent power, bandwidth, and safety constraints~\cite{even-chen_power-saving_2020}. In this context, spike-based neuromorphic processing provides an elegant compression mechanism: early-stage computation near the sensor reduces the data rate by encoding analog signals into sparse spike trains~\cite{basu_big_2017}, enabling more feasible wireless communication. Second, conventional architectures that separate memory and computing impose additional energy and latency overheads, as they buffer chunks of the data. Neuromorphic architectures inherently address this by co-locating memory and computation within neurons and synapses eliminating the need for additional processing stages. The challenges faced in cortical neural decoding closely parallel those emerging in motor unit-based \ac{EMG} decoding, where high-density recordings are increasingly adopted to resolve larger motor unit pools~\cite{farina_extraction_2025, varghese_design_2024}. As the volume and resolution of such signals grow, so too does the need for efficient, low-latency processing at the edge—making neuromorphic approaches a natural fit for this task.

Despite the conceptual compatibility between motor unit discharge activity and event-driven processing, neuromorphic approaches remain largely underexplored for proportional decoding based on motor unit activity. This gap arises in part from the challenge of effectively leveraging the temporal dynamics of biologically inspired neuron models. Membrane and synaptic time constants offer a powerful way to embed memory and computing within the same substrate. Yet, configuring or training these parameters remains non-trivial, particularly in settings with limited labeled data, calling for a combined strategy of manual tuning and data-driven optimization. As a result, most existing pipelines for proportional control using motor units sidestep this complexity by converting spike trains into analog firing rates through temporal binning and then applying conventional black-box machine learning approaches. While this workaround simplifies model design, it does not exploit the advantages of spike-based processing, it imposes a non-negligible latency, and increases memory overhead, during inference, making ultra-low power embedded electronic implementations unfeasible.

Here, we demonstrated that shallow \acp{SNN} with fixed synaptic time constants, can directly decode force trajectories from the decomposed motor unit spiking activity (Fig.\ref{fig:snn_to_lr_15_comparison}). With a modest number of trainable parameters—restricted to synaptic weights and membrane time constants—we achieved decoding performance comparable to standard linear regression baselines traditionally employed for motor decoding from motor units in isometric contractions~\cite{xu_estimation_2021}. Importantly, this was achieved without introducing explicit temporal binning or feature engineering stages, thus preserving the event-driven nature of the input data.

We further investigated how architectural design influences decoding robustness. Specifically, we demonstrated that inserting an intermediate spiking layer before the final continuous readout enhances resilience to decomposition noise, such as missing motor unit discharge times. Comparing both architectures under identical number of trainable parameters, we showed that networks incorporating a spiking intermediate layer exhibited greater tolerance to noise during inference (Fig.~\ref{fig:spike_omission}), owing to the thresholding mechanism of spiking neurons which suppresses minor input perturbations at the cost of a slight increased memory footprint (Fig.~\ref{fig:tradeoff_rmse_footprint}). 

Motor unit discharge activity provides a direct and physiologically grounded representation of the spinal output driving muscle contraction, and thus supports highly accurate proportional decoding. However, obtaining this information through decomposition remains computationally demanding. The process demands substantial computational time and memory resources, which scale with both the number of identified motor units and the number of \ac{EMG} channels, in order to reliably separate the discharge timings of individual motor units. For this reason, decomposition is typically performed offline, with separation matrices precomputed and later used for online decoding—an approach that, despite avoiding real-time source separation, still incurs non-negligible memory demands for storing and projecting signals through these matrices~\cite{rossato_i-spin_2024}. In light of these challenges, we explored an alternative strategy: encoding filtered \ac{HD-iEMG} signals directly into event-based spike trains using a layer of \ac{LIF} neurons. This method bypasses decomposition entirely, offering a lower-complexity alternative more suitable for real-time applications at the edge, albeit with an expected trade-off in decoding precision. We showed that shallow \acp{SNN} could still decode the gross structure of force trajectories from these encoded events with reasonable accuracy (Table~\ref{tab:summary_decoder_performance}). Decoding performance degraded compared to motor unit-based input, reflecting the expected information loss when moving from neural to global signal features~\cite{kapelner_predicting_2019}. Nevertheless, the performance gap (0.8–1.1 \%\ac{MVC}) suggests that event-based \ac{EMG} encoding could in practice be a viable compromise when decomposition is impractical or infeasible, particularly if combined with adaptive thresholding mechanisms~\cite{Sharifshazileh_Indiveri23} or heterogeneous neural populations to improve sensitivity to input amplitude variability~\cite{baracat_decoding_2024} without biasing the network training towards tasks with dense spike representations.

From a system-level perspective, our results support the development of a neuromorphic hardware implementation in which the first spiking layer operates near the acquisition unit to perform early-stage processing, either as an input encoding layer or as the first \ac{LIF} layer receiving event-based decomposed motor unit inputs. The second, continuous decoding layer could then be integrated into the end effector—for example, to generate real-time force commands (Fig.~\ref{fig:envisioned_pipeline}). Communication between these modules would rely on sparse spike trains, reducing the transmission data rate~\cite{biyan_combining_2025}. The vision is therefore to move towards a fully embedded multi-core neuromorphic solution for prosthesis control. Although such types of neuromorphic devices already exist~\cite{Richter_etal24,Greatorex_etal25}, they have been designed as general research platforms and do not yet include all the components identified in this study.

Our results provide specifications that can inform the development of a complete stand-alone hardware implementation. However, few aspects remain to be addressed in future work. The conclusions presented here are based on a limited sample size (two male participants), due to the practical constraints of intramuscular \ac{EMG} recordings. Larger and more diverse cohorts will be essential to improve statistical power and assess the robustness and generalizability of the decoding pipelines. Finally, the current implementation required training a separate decoder for each participant. This subject-specific approach was appropriate given the limited dataset, but is not scalable in practice. Future efforts could address this through hybrid training strategies, in which a general decoder is trained offline on pooled datasets and later adapted on-chip using lightweight calibration or few-shot learning to tailor performance to individual users. 

\section{Methods}\label{sec:methods}
\subsection{High-density intramuscular EMG dataset}\label{ssec:mdataset}
\ac{HD-iEMG} signals were recorded from two healthy male participants using three multi-channel intramuscular \acp{MEA}~\cite{muceli_accurate_2015} inserted into the forearm muscles. Electrode placement targeted the extrinsic muscles responsible for finger flexion and extension, and was guided by anatomical landmarks and real-time ultrasound imaging to ensure accurate and consistent positioning across both individuals. Each array consisted of 40 electrodes ($\SI{140}{\micro\meter} \times \SI{40}{\micro\meter}$), resulting in 120 recording sites in total. \ac{EMG} signals were acquired using a multichannel amplifier system (OT-Bioelettronica, Torino, Italy), sampled at \SI{10240}{\hertz}, and digitized at 16-bit resolution.

In parallel with the \ac{HD-iEMG} recordings, finger forces were measured using ten load cells (TAL220, \SI{10}{\kilogram} resolution), with two sensors assigned to each finger to independently capture flexion and extension force components. Further details regarding the electrode configuration and experimental setup are available in~\cite{grison_intramuscular_2024}.

Participants completed a series of isometric finger activation tasks, following a predefined trapezoidal force profile set at 15\% of their maximum voluntary contraction (\ac{MVC}). During each task, only one finger was actively engaged, while the remaining fingers remained at rest or producing minimal forces. Each force trajectory lasted \SI{26}{\second}, consisting of a 3-second ramp-up, a 20-second isometric hold, and a 3-second ramp-down. Tasks were performed sequentially, one finger and one direction (flexion or extension) at a time, and each task was repeated twice.

\subsection{Motor unit decomposition}\label{ssec:mdecomposition}
\ac{HD-iEMG} signals were decomposed into individual motor unit spike trains using a blind source separation algorithm designed for high-yield extraction~\cite{grison_particle_2024}. Decomposition was carried out independently for the isometric plateau phase of each finger task. For each task, signals from all \ac{EMG} electrodes and both repetitions were concatenated and jointly decomposed to extract task-specific motor unit activity.

Extracted motor units were post-processed to ensure their quality and physiological plausibility. Only motor units with a silhouette value exceeding 0.85 and an average discharge rate within the expected range for sustained isometric contractions (\SIrange{4}{50}{\hertz}) were retained. The silhouette score is a signal-based accuracy metric that measures the degree of separation between identified motor units and noise. Formally, it provides a normalized measure of clustering reliability, computed as the difference between within-cluster and between-cluster distances, applied to the separation between motor unit signals and noise~\cite{negro_multi-channel_2016}. Higher silhouette values indicate better separability and more reliable motor unit identification. To enhance temporal consistency and mitigate inter-repetition variability—particularly given the limited sample size—only motor units that were active in both repetitions of a given task were included in subsequent analyses. A detailed breakdown of the number of identified motor units units for each subject is presented in Supplementary Table~\ref{tabsupp:mu_decomposition}.

\subsection{Baseline regression models}\label{ssec:mbaseline}
For each movement direction (i.e., flexion, extension), decomposed motor unit spike trains from all corresponding finger tasks were pooled and segmented into \SI{80}{\milli\second} windows with 50\% overlap. The window length was selected based on a sweep analysis in which the baseline model was evaluated across multiple window sizes ranging from \SI{50}{\milli\second} to \SI{200}{\milli\second}. A window size of \SI{80}{\milli\second} provided the best trade-off between decoding performance and responsiveness and was therefore fixed for all subsequent analyses, including those involving the \ac{SNN} decoders. The degree of overlap between windows was not included in the sweep and was fixed to 50\%, in line with our previous work~\cite{grison_intramuscular_2024}.
Spike counts were computed within each window, min-max normalized, and used as input features to train a multi-output linear regression model. A linear decoder was chosen to provide a computationally efficient baseline and to mitigate overfitting given the limited dataset size.

Separate models were trained for each subject and movement direction to decode the continuous force trajectories of all five fingers from the binned motor unit discharge activity. This design choice was motivated by the fact that motor unit decomposition was performed separately for each direction, resulting in non-overlapping sets of identified units across flexion and extension. Enforcing motor unit tracking across both flexion and extension would have substantially reduced the number of usable units, thereby limiting the richness of the input signal. Instead, by training independent models for each direction, the decoder could leverage a larger and more representative motor unit population for each task. From an architectural standpoint, this corresponds to two non-overlapping weight sets that can be combined into a unified decoder, where weights for the opposite direction are set to zero. This formulation preserves modularity while enabling future extensions that integrate direction-specific decoding into a single, comprehensive model.

\subsection{Spiking neural network models}\label{ssec:msnn}
For each subject and movement direction, a single-layer \ac{SNN} was trained to estimate the continuous force trajectories of all five fingers. The network comprised \textit{$m$} input neurons, where \textit{$m$} corresponds to the total number of decomposed motor units pooled from all finger tasks associated with the given direction. Each input neuron projected to five output neurons, with each output dedicated to predicting the force produced by a specific finger.

\paragraph{Readout configurations}\label{par:msnn_readout}
We evaluated two configurations for the trainable readout layer: a \textit{leaky configuration} (\ac{LI} configuration) and a \textit{spiking configuration} (\ac{LIF} configuration) (Fig.~\ref{fig:envisioned_pipeline}). In the leaky configuration, output neurons were modeled as leaky integrators (\acp{LI}) that continuously accumulate synaptic input and directly produce smooth membrane potentials, which were later filtered to generate the final force predictions. In the spiking configuration, the output layer consisted of leaky integrate-and-fire (\ac{LIF}) neurons, which emit spikes when their membrane potentials exceed a threshold. The resulting spike trains were then passed to an intermediate layer of \ac{LI} neurons connected via instantaneous synapses with fixed weight, converting the discrete events into continuous-valued signals suitable for force decoding.

In both configurations, input motor unit spike trains were transmitted to the output neurons through synapses modeled as exponential kernels. Upon the arrival of a spike, a synaptic current $I_{syn}$ is generated and scaled by the corresponding synaptic weight. These currents decay exponentially over time with a fixed synaptic time constant $\tau_{syn}$ and are integrated into the membrane potential $U$ of each output neuron. Synaptic weights are optimized during training, while synaptic time constants remain fixed. The synaptic and membrane dynamics are described by the following equations:

\begin{align}
    I_{syn, j}[t+1] &= \alpha I_{syn, j}[t] + \sum_i W_{ji} \; x_i[t+1]\\
    U_j[t+1] &= \beta U_j[t] + I_{syn, j}[t+1]
\end{align}

\noindent
Here, $W_{ji}$ is the synaptic weight from input unit $MU_i$ to output neuron $j$, and $x_i[t]$ denotes the spike train of $MU_i$ at time $t$. $I_{syn, j}$ represents the cumulative synaptic input to neuron $j$, which is integrated into its membrane potential $U_j$. The decay factors $\alpha$ and $\beta$ are derived from the synaptic and membrane time constants, respectively, and are defined as $\alpha = e^{-\delta t / \tau_{syn}}$ and $\beta = e^{-\delta t / \tau_m}$, with a discrete simulation time step $\delta t = \SI{10}{\milli\second}$.

In the \ac{LIF} configuration, a firing threshold of \SI{0.045}{V} was applied to the output neurons. When this threshold was reached, the membrane potential was reset and a spike emitted. The resulting spike trains were first converted into continuous signals by an intermediate \ac{LI} layer with a fixed time constant of $\tau_{m,conv} = \SI{50}{\milli\second}$. As this conversion alone did not sufficiently suppress high-frequency fluctuations, a second \ac{LI} layer with a longer time constant of \SI{80}{\milli\second} was added—a double-stage filtering step applied only in the \ac{LIF} configuration (Fig.~\ref{fig:mu_versus_count_schematic}).
In both configurations, the final continuous force estimates were then obtained by low-pass filtering the membrane potentials, following the same procedure as in the baseline model (see Methods~\ref{ssec:mbaseline}). 

All \ac{SNN} models were implemented using \textit{SnnTorch}~\cite{eshraghian_training_2023}, and simulations were performed with a discrete time step of $\Delta t = \SI{10}{\milli\second}$. \ac{SNN} hyper-parameters are summarized in Table~\ref{tab:snn_parameters}.

For the \ac{LI} network configurations, the membrane time constants ($\tau_m$) of the output neurons in the first \ac{LI} layer was fixed to \SI{80}{\milli\second}. For the spiking configurations, given the bigger parameter space due to the added intermediate \ac{LIF} layer, we initialized the readout neurons time constants to a normal distribution with mean \SI{30}{\milli\second} and a standard deviation of \SI{1.5}{\milli\second} (5\% of the mean). These values were treated as additional learnable parameters and jointly optimized with the synaptic weights during training.

\begin{table}[ht]
\centering
\caption{Summary of the \acp{SNN} hyper-parameters.}
\begin{tabular}{>{\centering\arraybackslash}p{2.1cm}|p{7.1cm}|p{4.6cm}}
\toprule
\textbf{Parameter} & \textbf{Description} & \textbf{Value} \\
\midrule
$\Delta t$ & Simulation time step & \SI{10}{\milli\second} \\
$\tau_{syn}$ & Fixed synaptic time constant (exponential kernel) & \SI{10}{\milli\second} \\
\multirow[t]{2}{*}{weights} & \multirow[t]{2}{*}{Synaptic weight initialization range} & \ac{LI} configuration: $U(-0.1,\,0.1)$ \\
& & \ac{LIF} configuration: $U(-0.01,\,0.01)$ \\
\addlinespace
\multirow[t]{2}{*}{$\tau_{m}$} & \multirow[t]{2}{7.2cm}{Membrane time constant of first layer of neurons (fixed or trainable)} & \ac{LI} configuration: \SI{30}{\milli\second} \\
& & \ac{LIF} configuration: $\mathcal{N}(\SI{30}{\milli\second},\SI{1.5}{\milli\second})$ \\
\addlinespace
$\tau_{m,conv}$ & Time constant for spike-to-continuous conversion layer & \ac{LIF} configuration: \SI{50}{\milli\second} \\
threshold & Firing threshold for \ac{LIF} output neurons& \ac{LIF} configuration \SI{0.045}{V} \\
Train batch size & Number of samples per training batch & 16 \\
Test batch size & Number of samples per evaluation batch & 1 \\
Epochs & Number of training epochs & 100 \\
\multirow[t]{2}{*}{Learning rate} & \multirow[t]{2}{*}{Optimizer learning rate} & \ac{LI} configuration: 0.01 \\
& & \ac{LIF} configuration: 0.001 \\
\addlinespace
\bottomrule
\end{tabular}
\label{tab:snn_parameters}
\end{table}

\paragraph{Training and inference}\label{par:mtraining}
All \ac{SNN} models were trained using surrogate gradient descent with a mean squared error loss and Adam optimizer. During training, the input repetition was segmented into overlapping 10-second windows with a 50\% overlap. This window length was selected to balance computational efficiency and temporal context, ensuring that each training update was informed by a sufficient duration of motor unit activity  and force output.

Once training was complete, the network was evaluated in inference mode under conditions that approximate real-time deployment, such as on always-on neuromorphic hardware. During inference, the batch size was set to one, and the model processed 10-second spike train segments sequentially. At each simulation time step ($\Delta t = \SI{10}{\milli\second}$), it produced a continuous five-dimensional force estimate. To simulate uninterrupted, stateful processing within the PyTorch framework, neuron and synapse states were preserved across the segments without resets between inference windows. This workaround was applied solely to ensure a fair comparison with the baseline model. In mixed analog neuromorphic hardware, such state management is inherently handled, requiring no explicit intervention or additional memory buffers.

\subsection{Evaluation metrics}\label{ssec:mmetrics}
Model performance was assessed using standard regression metrics: \ac{RMSE}, \ac{MAE}, and the \ac{R2}, computed under a 2-fold cross-validation scheme. Among these, \ac{RMSE} was selected as the primary evaluation metric in the main text due to its interpretability: it quantifies the deviation from the target force as a percentage of \ac{MVC}, providing a concrete measure of decoding accuracy.

To mitigate the risk of overfitting to fixed activation sequences, the order of finger activations was shuffled within each repetition prior to training. This randomized ordering was applied consistently across all models to ensure a fair comparison.

\subsection{SNN robustness analysis}\label{ssec:mrobustness}
To evaluate the robustness of the two \ac{SNN} configurations to noise introduced at inference, networks were trained on clean decomposed motor unit spike trains and subsequently tested with varying levels of spike omission. This omission simulates decomposition-related noise that may occur during real-time applications, where some motor unit discharge times may go undetected. Performance was assessed using \ac{RMSE} across tasks and movement directions, as spike omission rates ranged from 0\% to 50\% with 10\% increments.

\subsection{EMG-to-spike encoding}\label{ssec:mencoding}
A layer of \ac{LIF} neurons was used to convert the continuous \ac{iEMG} signals into an event-based representation compatible with \ac{SNN} processing. Each of the 120 \ac{iEMG} recording channels—acquired from three distinct intramuscular microelectrode arrays—was assigned to a dedicated \ac{LIF} neuron. The filtered \ac{iEMG} signals were injected as continuous input currents into the respective encoding neurons.

Given that the three arrays were implanted at anatomically distinct sites, the amplitude distributions of the recorded \ac{iEMG} signals differed across electrodes—a difference that was particularly pronounced in Subject~1. To account for this variability, the parameters of the encoding neurons were adjusted on a per-electrode basis. Specifically, all 40 encoding neurons associated with a given electrode shared a common membrane time constant, $\tau_{\text{enc},m}$, and firing threshold. A homogeneous membrane time constant of \SI{20}{\milli\second} was used across all neurons. In contrast, the firing thresholds were adapted to the signal characteristics of each electrode: \SI{0.1}{\volt} for electrode 1, \SI{0.4}{\volt} for electrode 2, and \SI{0.2}{\volt} for electrode 3. For Subject~2, where inter-electrode variability was less pronounced, fixed thresholds and membrane time constants of \SI{0.06}{\volt} and \SI{20}{\milli\second} were applied uniformly across all electrodes.

The choice of thresholds was guided by the resulting event rates of the encoding layer. Since the encoded spike trains were streamed directly into the same \ac{SNN} backend used in the motor unit decoding experiments, the tuning aimed to match the event rate regime observed for the decomposed motor units. This ensured compatibility with the downstream network and avoided the need for re-calibration of the readout layer parameters.

\subsection{Decoder memory footprint}\label{ssec:mfootprint}
The memory footprint of each decoder was estimated based on the number of parameters and their storage precision. All parameters in the baseline linear regression models and \ac{SNN} decoders were stored at full precision (4 bytes). Only nonzero parameters were included in the calculation, excluding pruned (zero-valued) weights. In the case of one-to-one channel encoding, only the diagonal elements of the weight matrix were counted, as they represent the actual connections used. This approach provides a more accurate reflection of the memory requirements during inference. The estimate follows the static formulation proposed in~\cite{Yik_etal25} and excludes additional runtime memory.

For all decoders operating on motor unit spike trains, we further included the memory required to store the separation matrix used during decomposition. This matrix, composed of 32-bit floats, has dimensions equal to the number of identified motor units multiplied by the number of \ac{EMG} channels.

\subsection{Statistical analysis}
We used a two-tailed Mann-Whitney U test to compare the baseline model (\ac{LR}) with the \ac{SNN} models (see Fig.~\ref{fig:boxplot_comparison}). Each group contained $n=20$ sample, corresponding to the \ac{RMSE} values obtained across the two repetitions, five finger tasks and the two directions of movement.
\vspace{.8cm}

\bmhead{Acknowledgements}
FB was supported by the Candoc grant (project NeuroDEX grant number FK-25-080) from the University of Zurich and  PhD grant 23 Neuroscience Center Zurich (ZNZ). AG was supported by UK Research and Innovation (UKRI Centre for Doctoral Training in AI for Healthcare grant number EP/S023283/1) and by Huawei Technologies Research \& Development (UK) Limited. DF was supported by NaturalBionicS (ERC Synergy 810346) and NISNEM (EPSRC
EP/T020970/1).

\bmhead{Data availability} Data will be made available from the corresponding author (FB) upon reasonable request.

\bmhead{Code availability} The code developed to analyse the data is available from the corresponding author (FB) on request.

\bmhead{Author contribution}
FB, ED conceptualized the study. AG provided the recordings, and decomposed motor units. FB conducted simulations, analyzed, visualized, and interpreted the data, and wrote the original draft. FB, AG, DF, GI, ED edited the manuscript for important scientific content and all approved the final version.

\bmhead{Competing interests}
The authors declare that there are no competing interests.

\clearpage
\begin{appendices}
\section*{Supplementary Material}

\begin{table}[h!]
\centering
\caption{Summary of baseline and \ac{SNN} decoders' performance. Metrics are shown separately for extension and flexion, with bold entries representing the average across both directions. Results are averaged across 10 different initialization seeds chosen at random for the SNN decoders.}
\begin{tabular}{c|c|c|c|c|c|c}
\toprule
\textbf{Subject} & \textbf{Input Type} & \textbf{Decoder Model} & \textbf{Direction} & \textbf{RMSE} (\%\ac{MVC}) & \textbf{MAE} (\%\ac{MVC}) & \textbf{R$^2$} \\
\midrule
\multirow{12}{*}{S1}
  & \multirow{3}{*}{motor unit spike count}
  & \multirow{3}{*}{Baseline (LR)}
  & Flexion    & $1.31 \pm 0.16$ & $0.77 \pm 0.08$ & $0.94 \pm 0.02$ \\
  &  &  & Extension  & $2.16 \pm 0.42$ & $1.24 \pm 0.19$ & $0.83 \pm 0.07$ \\
  &  &  & \textbf{Avg.}      & $\mathbf{1.73 \pm 0.29}$ & $\mathbf{1.01 \pm 0.14}$ & $\mathbf{0.88 \pm 0.04}$ \\
  \cmidrule{2-7}
  & \multirow{6}{*}{motor unit spike train} 
  & \multirow{3}{*}{\ac{SNN} (LI)} 
  & Flexion    & $1.50 \pm 0.30$ & $0.96 \pm 0.17$ & $0.92 \pm 0.03$ \\
  &  &  & Extension  & $1.96 \pm 0.16$ & $1.22 \pm 0.08$ & $0.87 \pm 0.02$ \\
  &  &  & \textbf{Avg.}      & $\mathbf{1.73 \pm 0.23}$ & $\mathbf{1.09 \pm 0.13}$ & $\mathbf{0.89 \pm 0.03}$ \\
  \cmidrule{3-7}
  &  & \multirow{3}{*}{\ac{SNN} (LIF)} 
  & Flexion    & $1.48 \pm 0.22$ & $0.95 \pm 0.20$ & $0.92 \pm 0.03$ \\
  &  &  & Extension  & $2.08 \pm 0.33$ & $1.25 \pm 0.22$ & $0.85 \pm 0.05$ \\
  &  &  & \textbf{Avg.}      & $\mathbf{1.78 \pm 0.27}$ & $\mathbf{1.10 \pm 0.21}$ & $\mathbf{0.88 \pm 0.04}$ \\
  \cmidrule{2-7}
  & \multirow{3}{*}{encoded \ac{HD-iEMG}} 
  & \multirow{3}{*}{\ac{SNN} (LI)}
 & Flexion  & $2.98 \pm 1.52$ & $1.98 \pm 1.04$ & $0.60 \pm 0.32$ \\
  &    &    & Extension  & $2.13 \pm 0.75$ & $1.45 \pm 0.46$ & $0.82 \pm 0.12$ \\
  &    &    & \textbf{Avg.} & $\mathbf{2.55 \pm 1.13}$ & $\mathbf{1.72 \pm 0.75}$ & $\mathbf{0.71 \pm 0.22}$ \\
\midrule
\multirow{9}{*}{S2}
  & \multirow{3}{*}{motor unit spike count}
  & \multirow{3}{*}{Baseline (LR)}
  & Flexion    & $2.21 \pm 0.35$ & $1.35 \pm 0.17$ & $0.82 \pm 0.05$ \\
  &  &  & Extension  & $2.06 \pm 0.29$ & $1.37 \pm 0.15$ & $0.86 \pm 0.04$ \\
  &  &  & \textbf{Avg.}      & $\mathbf{2.13 \pm 0.32}$ & $\mathbf{1.36 \pm 0.16}$ & $\mathbf{0.84 \pm 0.05}$ \\
  \cmidrule{2-7}
  & \multirow{6}{*}{motor unit spike train} 
  & \multirow{3}{*}{\ac{SNN} (LI)} 
  & Flexion    & $2.47 \pm 0.48$ & $1.71 \pm 0.40$ & $0.79 \pm 0.07$ \\
  &  &  & Extension  & $2.16 \pm 0.19$ & $1.57 \pm 0.17$ & $0.86 \pm 0.03$ \\
  &  &  & \textbf{Avg.}      & $\mathbf{2.31 \pm 0.34}$ & $\mathbf{1.64 \pm 0.29}$ & $\mathbf{0.82 \pm 0.05}$ \\
  \cmidrule{3-7}
  &  & \multirow{3}{*}{\ac{SNN} (LIF)} 
  & Flexion    & $2.25 \pm 0.60$ & $1.50 \pm 0.56$ & $0.82 \pm 0.09$ \\
  &  &  & Extension  & $1.99 \pm 0.21$ & $1.41 \pm 0.19$ & $0.88 \pm 0.03$ \\
  &  &  & \textbf{Avg.}      & $\mathbf{2.12 \pm 0.41}$ & $\mathbf{1.46 \pm 0.38}$ & $\mathbf{0.85 \pm 0.06}$ \\
  \cmidrule{2-7}
  & \multirow{3}{*}{encoded \ac{HD-iEMG}} 
  & \multirow{3}{*}{\ac{SNN} (LI)}
  & Flexion & $3.55 \pm 0.93$ & $2.41 \pm 0.36$ & $0.55 \pm 0.25$ \\
  & &  & Extension  & $3.28 \pm 1.54$ & $2.35 \pm 0.98$ & $0.59 \pm 0.39$ \\
  &  &   & \textbf{Avg.}& $\mathbf{3.41 \pm 1.23}$ & $\mathbf{2.38 \pm 0.67}$ & $\mathbf{0.57 \pm 0.32}$ \\
\bottomrule
\end{tabular}
\label{tabsupp:mu_directional_performance_full}
\end{table}

\newpage

\begin{table}[h!]
\centering
\caption{Average test performance metrics for each decoder model under increasing spike omission levels. Values are reported as means with standard deviations, computed across finger tasks and movement directions.}
\begin{tabular}{c|c|c|c|c|c}
\toprule
\textbf{Subject}  & \textbf{Decoder Model} & \textbf{Omission Level (\%)} & \textbf{RMSE} (\%MVC) & \textbf{MAE} (\%MVC) & \textbf{R$^2$} \\
\midrule
\multirow{18}{*}{S1}
  & \multirow{7}{*}{Baseline (LR)} & \textbf{0}  & $\mathbf{1.73 \pm 0.26}$ & $\mathbf{1.01 \pm 0.18}$ & $\mathbf{0.88 \pm 0.04}$ \\
  &                                           & 10 & $1.90 \pm 0.29$ & $1.11 \pm 0.22$ & $0.86 \pm 0.05$ \\
  &                                           & 20 & $2.20 \pm 0.32$ & $1.28 \pm 0.23$ & $0.82 \pm 0.05$ \\
  &                                           & 30 & $2.60 \pm 0.32$ & $1.48 \pm 0.24$ & $0.76 \pm 0.06$ \\
  &                                           & 40 & $3.04 \pm 0.31$ & $1.70 \pm 0.26$ & $0.67 \pm 0.07$ \\
  &                                           & 50 & $3.53 \pm 0.29$ & $1.93 \pm 0.29$ & $0.56 \pm 0.06$ \\
\cmidrule{2-6}
  & \multirow{7}{*}{\ac{SNN} \ac{LI}}    & \textbf{0}  & $\mathbf{1.74 \pm 0.32}$ & $\mathbf{1.10 \pm 0.17}$ & $\mathbf{0.89 \pm 0.04}$ \\
  &                                           & 10 & $1.84 \pm 0.33$ & $1.18 \pm 0.21$ & $0.88 \pm 0.04$ \\
  &                                           & 20 & $2.07 \pm 0.33$ & $1.34 \pm 0.25$ & $0.85 \pm 0.04$ \\
  &                                           & 30 & $2.40 \pm 0.27$ & $1.54 \pm 0.26$ & $0.80 \pm 0.04$ \\
  &                                           & 40 & $2.79 \pm 0.23$ & $1.77 \pm 0.29$ & $0.74 \pm 0.04$ \\
  &                                           & 50 & $3.22 \pm 0.22$ & $2.01 \pm 0.31$ & $0.65 \pm 0.04$ \\
\cmidrule{2-6}
  & \multirow{7}{*}{\ac{SNN} \ac{LIF}}  & \textbf{0}  & $\mathbf{1.84 \pm 0.31}$ & $\mathbf{1.13 \pm 0.25}$ & $\mathbf{0.88 \pm 0.05}$ \\
  &                                           & 10 & $1.88 \pm 0.34$ & $1.18 \pm 0.26$ & $0.87 \pm 0.05$ \\
  &                                           & 20 & $1.98 \pm 0.35$ & $1.26 \pm 0.28$ & $0.86 \pm 0.05$ \\
  &                                           & 30 & $2.16 \pm 0.37$ & $1.39 \pm 0.28$ & $0.83 \pm 0.06$ \\
  &                                           & 40 & $2.43 \pm 0.36$ & $1.55 \pm 0.29$ & $0.79 \pm 0.07$ \\
  &                                           & 50 & $2.81 \pm 0.40$ & $1.76 \pm 0.29$ & $0.72 \pm 0.08$ \\
\midrule
\multirow{18}{*}{S2}
  & \multirow{7}{*}{Baseline (LR)} & \textbf{0}  & $\mathbf{2.13 \pm 0.54}$ & $\mathbf{1.36 \pm 0.45}$ & $\mathbf{0.84 \pm 0.08}$ \\
  &                                           & 10 & $2.29 \pm 0.57$ & $1.47 \pm 0.50$ & $0.81 \pm 0.09$ \\
  &                                           & 20 & $2.56 \pm 0.57$ & $1.65 \pm 0.53$ & $0.77 \pm 0.09$ \\
  &                                           & 30 & $2.94 \pm 0.55$ & $1.86 \pm 0.53$ & $0.71 \pm 0.10$ \\
  &                                           & 40 & $3.38 \pm 0.52$ & $2.11 \pm 0.56$ & $0.62 \pm 0.11$ \\
  &                                           & 50 & $3.85 \pm 0.48$ & $2.36 \pm 0.59$ & $0.51 \pm 0.11$ \\
\cmidrule{2-6}
  & \multirow{7}{*}{\ac{SNN} \ac{LI}}  & \textbf{0}  & $\mathbf{2.32 \pm 0.46}$ & $\mathbf{1.64 \pm 0.40}$ & $\mathbf{0.82 \pm 0.06}$ \\
  &                                           & 10 & $2.41 \pm 0.54$ & $1.73 \pm 0.40$ & $0.81 \pm 0.08$ \\
  &                                           & 20 & $2.59 \pm 0.57$ & $1.88 \pm 0.42$ & $0.78 \pm 0.08$ \\
  &                                           & 30 & $2.85 \pm 0.54$ & $2.07 \pm 0.43$ & $0.74 \pm 0.09$ \\
  &                                           & 40 & $3.16 \pm 0.48$ & $2.27 \pm 0.45$ & $0.68 \pm 0.09$ \\
  &                                           & 50 & $3.52 \pm 0.42$ & $2.50 \pm 0.47$ & $0.60 \pm 0.08$ \\
\cmidrule{2-6}
  & \multirow{7}{*}{\ac{SNN} \ac{LIF}}    & \textbf{0}  & $\mathbf{2.08 \pm 0.54}$ & $\mathbf{1.42 \pm 0.49}$ & $\mathbf{0.85 \pm 0.07}$ \\
  &                                           & 10 & $2.14 \pm 0.55$ & $1.48 \pm 0.51$ & $0.85 \pm 0.08$ \\
  &                                           & 20 & $2.26 \pm 0.55$ & $1.58 \pm 0.52$ & $0.83 \pm 0.08$ \\
  &                                           & 30 & $2.46 \pm 0.58$ & $1.73 \pm 0.56$ & $0.80 \pm 0.09$ \\
  &                                           & 40 & $2.74 \pm 0.55$ & $1.92 \pm 0.57$ & $0.75 \pm 0.10$ \\
  &                                           & 50 & $3.11 \pm 0.55$ & $2.15 \pm 0.58$ & $0.68 \pm 0.11$ \\
\bottomrule
\end{tabular}
\label{tabsupp:spike_omission_robustness}
\end{table}

\newpage

\begin{table}[ht]
\centering
\caption{Motor unit decomposition and encoding statistics at 15\% \ac{MVC}. Discharge rates of motor units and encoded iEMG channels are computed across both repetitions and finger tasks for each subject.}
\label{tabsupp:mu_decomposition}
\begin{tabular}{>{\centering\arraybackslash}p{1cm} >{\centering\arraybackslash}p{2cm} >{\centering\arraybackslash}p{2cm} >{\centering\arraybackslash}p{2cm}  >
{\centering\arraybackslash}p{3.2cm} >
{\centering\arraybackslash}p{3.2cm}}
\toprule
\textbf{Subject} & \textbf{Direction} & \textbf{Number of motor units} &  \textbf{Number of iEMG channels} &\textbf{Mean motor units discharge rate (pps)} & \textbf{Mean firing rate of encoded channels (Hz)} \\
\midrule
S1 & Flexion   & 121 & 120 & $11.90 \pm 3.70$ & $4.62 \pm 0.88$ \\
   & Extension & 196 & 120 & $13.00 \pm 4.72$ & $6.62 \pm 0.50$ \\
\midrule
S2 & Flexion   &  51 & 120 & $10.78 \pm 3.05$ & $11.57 \pm 1.04$ \\
   & Extension &  93 & 120 & $12.66 \pm 4.38$ & $14.69 \pm 0.72$ \\
\bottomrule 
\end{tabular}
\end{table}

\begin{table}[ht!]
\renewcommand{\arraystretch}{1.4}
\centering
\caption{Static metrics for decoder models. Reported parameter counts include the linear coefficients for the baseline model, and the synaptic weights, biases, and time constants for neurons and synapses in the \ac{SNN} architectures. The encoding \ac{SNN} receives spike-encoded \ac{EMG} channel inputs and maintains a consistent footprint across movement directions.}
\label{tabsupp:static_metrics}
\begin{tabular}{
>{\centering\arraybackslash}m{0.7cm} 
>{\centering\arraybackslash}m{2.2cm} 
>{\centering\arraybackslash}m{1.1cm} 
>{\centering\arraybackslash}m{1.8cm} 
>{\centering\arraybackslash}m{1.6cm} 
>{\centering\arraybackslash}m{2.5cm} 
>{\centering\arraybackslash}m{2.5cm} 
>{\centering\arraybackslash}m{0.8cm}}
\toprule
\textbf{Subject} & \textbf{Decoder Model} & \textbf{Direction} & \textbf{Total Parameters} & \textbf{Trainable Parameters} & \textbf{Parameters Memory (Bytes)} & \textbf{Decomposition Memory (Bytes)} & \textbf{Latency (ms)} \\
\midrule
\multirow{7}{*}{S1} 
  & Baseline (\ac{LR}) & Flex.   & 605   & 605   & 2420  & 58080  & 40 \\
  &                    & Ext.    & 980   & 980   & 3920  & 94080  & 40 \\\cmidrule{2-8}
  & SNN (\ac{LI})      & Flex.   & 615   & 610   & 2456  & 58080  & 10 \\
  &                    & Ext.    & 990   & 985   & 3956  & 94080  & 10 \\\cmidrule{2-8}
  & SNN (\ac{LIF})     & Flex.   & 657   & 615   & 2612  & 58080  & 10 \\
  &                    & Ext.    & 1032  & 990   & 4112  & 94080  & 10 \\\cmidrule{2-8}
  & Encoding SNN (\ac{LI}) & Flex./Ext. & 15256 & 605 & 4848  & 0  & 10 \\
\midrule
\multirow{7}{*}{S2} 
  & Baseline (\ac{LR}) & Flex.   & 255   & 255   & 1020  & 24480  & 40 \\
  &                    & Ext.    & 465   & 465   & 1860  & 44640  & 40 \\\cmidrule{2-8}
  & SNN (\ac{LI})      & Flex.   & 265   & 260   & 1056  & 24480  & 10 \\
  &                    & Ext.    & 475   & 470   & 1896  & 44640  & 10 \\\cmidrule{2-8}
  & SNN (\ac{LIF})     & Flex.   & 307   & 265   & 1212  & 24480  & 10 \\
  &                    & Ext.    & 517   & 475   & 2052  & 44640  & 10 \\\cmidrule{2-8}
  & Encoding SNN (\ac{LI}) & Flex./Ext. & 15256 & 605 & 4848  & 0  & 10 \\
\bottomrule
\end{tabular}
\end{table}

\clearpage
\subsection*{Force estimation across models for Subject~1 during extension tasks}
\begin{figure}[h!]
    \centering
    \includegraphics[width=1\linewidth]{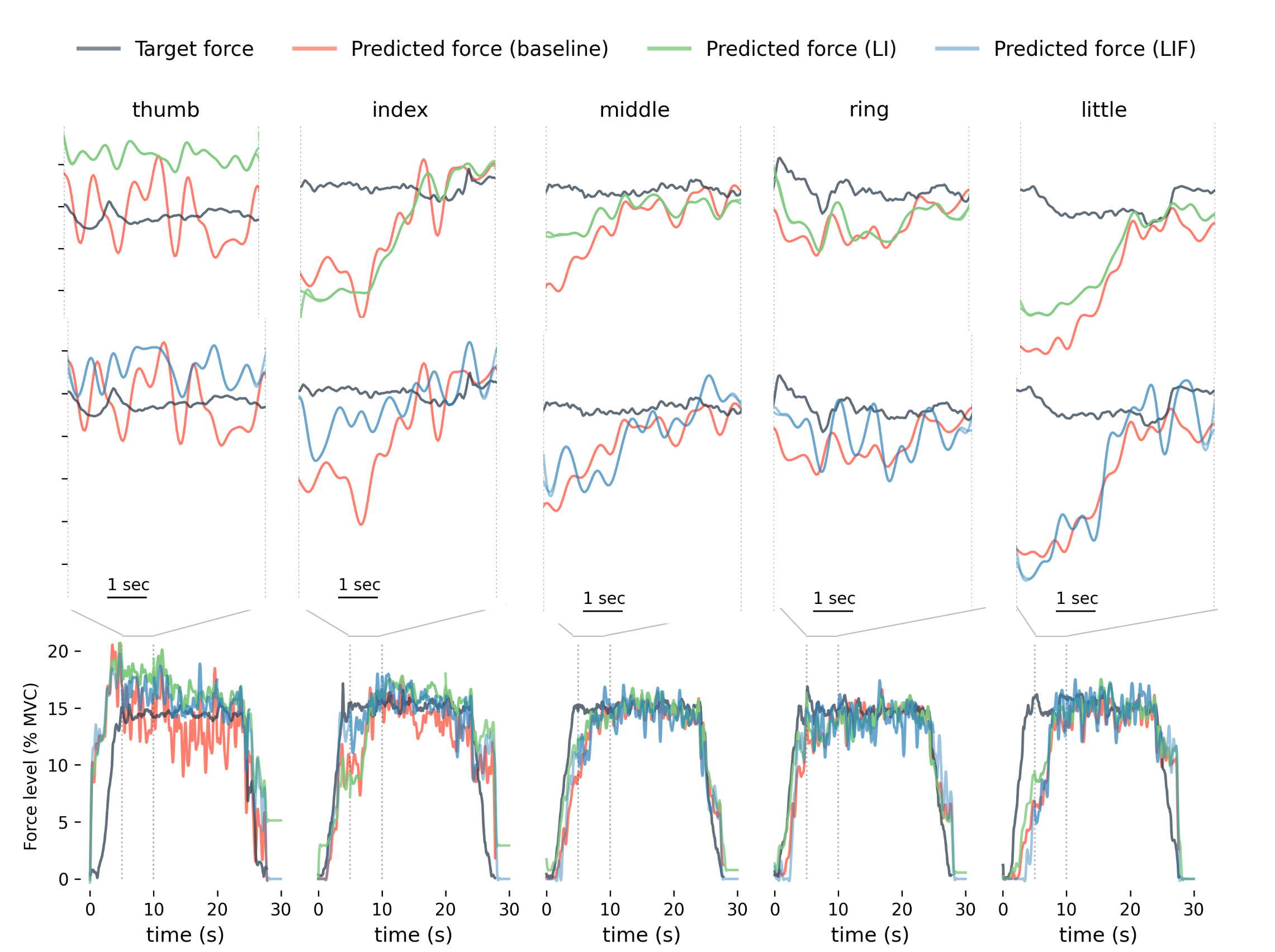}
    \caption{Example force estimation traces for Subject~1 during individual finger extension tasks (repetition~2), using motor unit spike trains as inputs, with zoomed insets highlighting the interval between 5 and 10 seconds. Only the predicted forces for the active (targeted) finger are shown for clarity.}
    \label{figsupp:predictions_comparison_s1_ext_rep2}
\end{figure}

\newpage
\subsection*{Force estimation across models for Subject~2 during extension and flexion tasks}
\begin{figure}[H]
    \centering
    \includegraphics[width=1\linewidth]{
     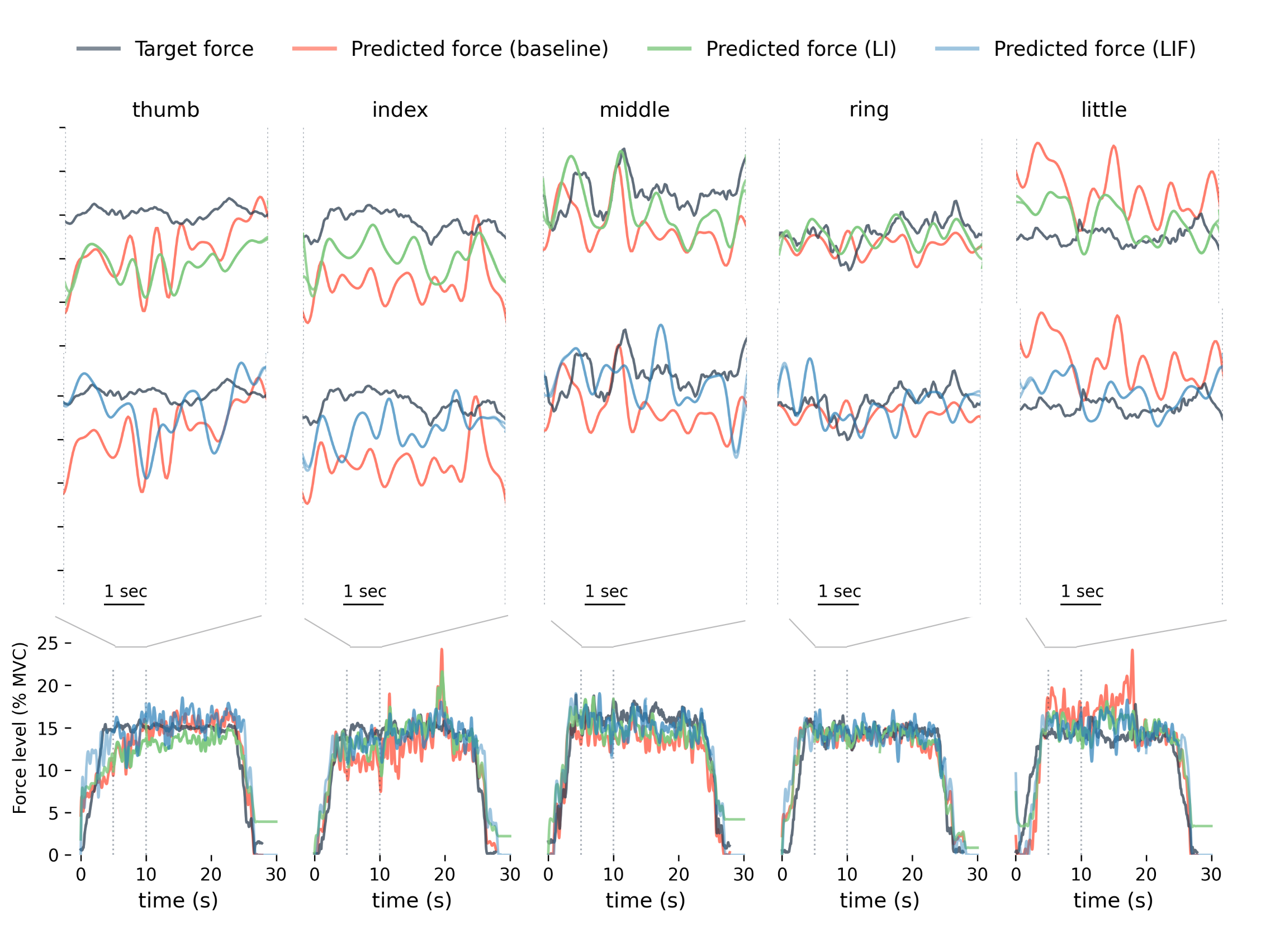}
    \caption{Example force estimation traces for Subject~2 during individual finger extension tasks (repetition~2), using motor unit spike trains as inputs, with zoomed insets highlighting the interval between 5 and 10 seconds. Only the predicted forces for the active (targeted) finger are shown for clarity.}
    \label{figsupp:predictions_comparison_s2_ext_rep2}
\end{figure}

\begin{figure}[H]
    \centering
    \includegraphics[width=1\linewidth]{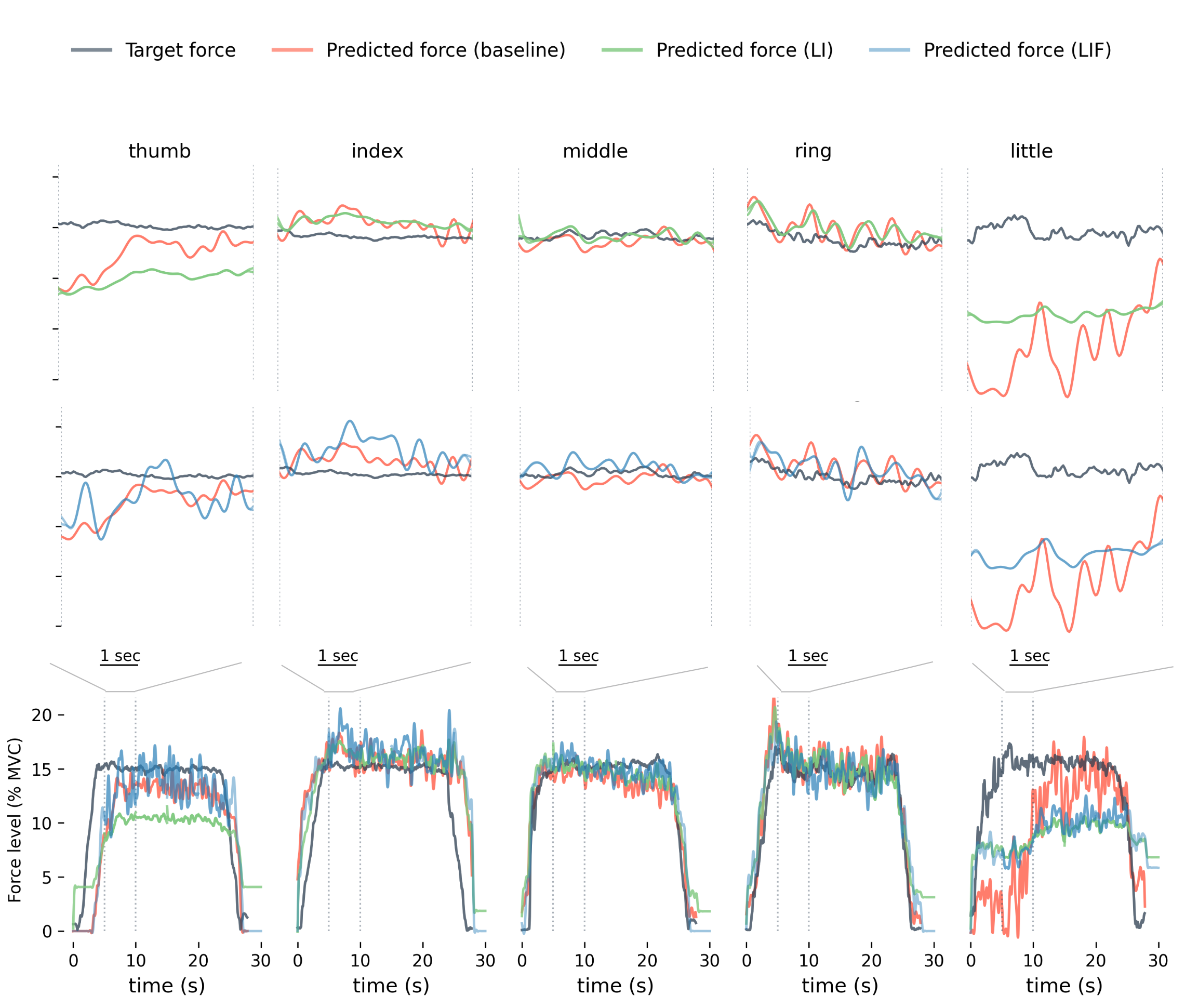}
    \caption{Example force estimation traces for Subject~2 during individual finger flexion tasks (repetition~2), using motor unit spike trains as inputs, with zoomed insets highlighting the interval between 5 and 10 seconds. Only the predicted forces for the active (targeted) finger are shown for clarity.}
    \label{figsupp:predictions_comparison_s2_flex_rep2}
\end{figure}

\clearpage
\subsection*{Relation of encoded \ac{HD-iEMG} to global features}
\begin{figure}[h!]
\begin{subfigure}[]{0.9\linewidth}
    {\includegraphics[width=1\linewidth]
{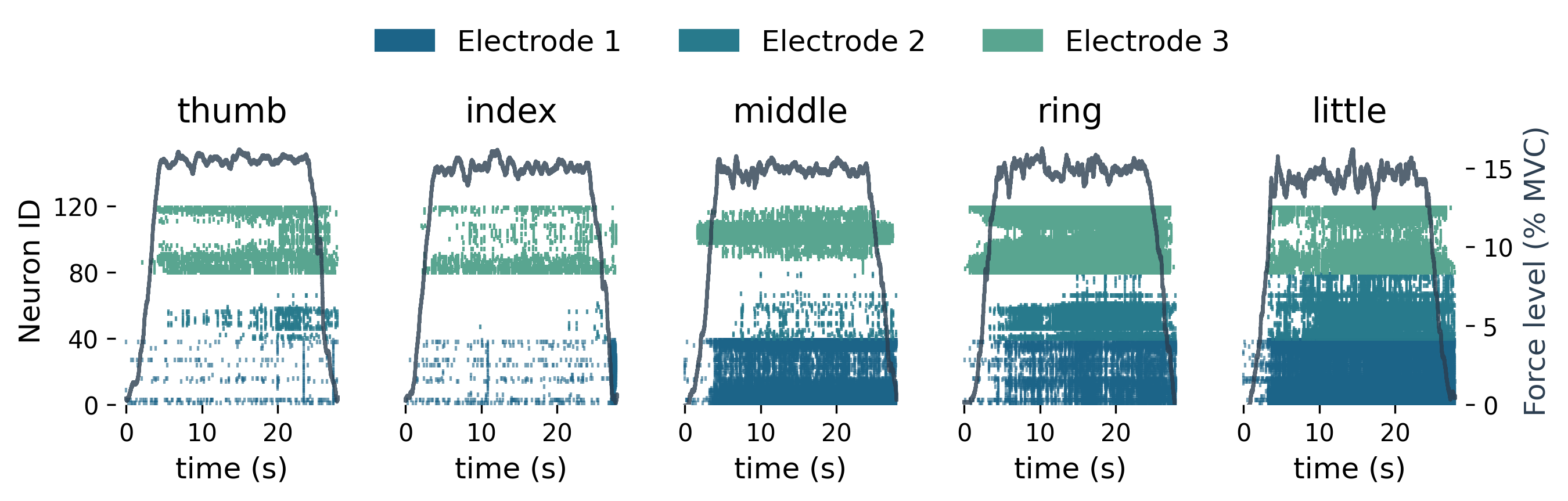}}
    \caption{}
    \label{figsupp:example_encoding_flexion}
\end{subfigure}
\begin{subfigure}[]{0.9\linewidth}
     {\includegraphics[width=1\linewidth]{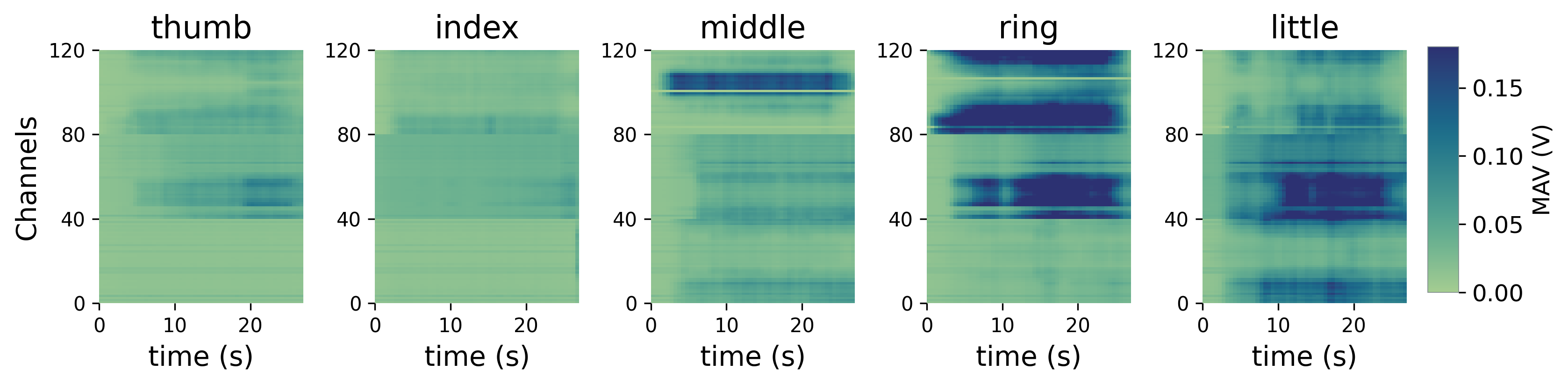}}
    \caption{}
    \label{figsupp:input_emg_flexion}
\end{subfigure}
\caption{Example of spike-encoded representations and corresponding \ac{iEMG} during individual finger flexion tasks (Subject~1, repetition~2).
(a) Raster plots of spikes generated by \ac{LIF} neurons in response to the \ac{iEMG} inputs. Force trajectories (in \%\ac{MVC}) are overlaid for reference. Neurons are color-coded by electrode. Spike encoding was performed using \ac{LIF} neurons with a homogeneous membrane time constant of \SI{20}{\milli\second}, while thresholds were adjusted per electrode to account for amplitude differences (Electrode 1: \SI{0.1}{\volt}, Electrode 2: \SI{0.4}{\volt}, Electrode 3: \SI{0.2}{\volt}).
(b) Corresponding input \ac{iEMG} signals for each finger task smoothened with a \ac{MAV} on windows of 1-second and 50\% overlap. Each column represents one finger task, and all channels from the three electrodes (120 total) are included.}
\label{figsupp:encoding_input_emg_flexion}
\end{figure}

\end{appendices}

\clearpage
\printbibliography
\end{document}